\documentclass[aps, rmp, showpacs, nofootinbib,10pt, longbibliography,twocolumn, amsfonts, amsmath, amssymb, superscriptaddress, floatfix]{revtex4-1}
\usepackage[colorlinks,bookmarks=false,citecolor=blue,linkcolor=blue,urlcolor=black]{hyperref}

\usepackage[T1]{fontenc}
\usepackage[latin9]{inputenc}
\usepackage{color}
\usepackage{graphicx}
\usepackage{epsfig}
\usepackage{ulem}
\def\be{\begin{equation}}
\def\ee{\end{equation}}
\def\bea{\begin{eqnarray}}
\def\eea{\end{eqnarray}}
\def\ra{\rangle}
\def\la{\langle}

\def\bi{\begin{itemize}}
\def\ei{\end{itemize}}

\newcommand{\vect}[1]{\mathbf{#1}}

\begin{document}

\title{Time crystals: a review}

\author{Krzysztof Sacha and Jakub Zakrzewski} 
\affiliation{
Instytut Fizyki imienia Mariana Smoluchowskiego and  Mark Kac Complex Systems Research Center, Uniwersytet Jagiello\'nski ulica Profesora Stanis\l{}awa \L{}ojasiewicza 11 PL-30-348 Krak\'ow Poland}
 
\pacs{11.30.Qc, 67.80.-s, 73.23.-b, 03.75.Lm, 67.85.Hj, 67.85.Jk}
\date{\today}
\begin{abstract}
Time crystals are time-periodic self-organized structures postulated by  Frank Wilczek in 2012.
While the original concept was strongly criticized, it stimulated at the same time an intensive
research leading to propositions and experimental verifications of discrete (or Floquet) time crystals
-- the structures that appear in the time domain due to spontaneous breaking of discrete time translation symmetry.
The struggle to observe discrete time crystals is reviewed here together with propositions that generalize this concept
introducing condensed matter like physics in the time domain.
We shall also revisit the original Wilczek's idea and review strategies aimed at spontaneous breaking of continuous time translation symmetry.
\end{abstract}

\maketitle
\tableofcontents

\section{Introduction}
Crystals are everywhere ranging from jewelery to salt in the kitchen. They are built of atoms in a regular arrangement in space and  they  have distinct mechanical properties as well as  heat or electric conductance properties. Crystals are formed due to mutual interactions between atoms that self-organize and build regular structures in space. This kind of self-organization is a quantum mechanical phenomenon and is related to spontaneous space translation symmetry breaking \cite{Strocchi2005}. It is often neglected that {a probability density for measurement of a position of a single particle in a system of} mutually interacting  {particles} cannot reveal a crystalline structure unless continuous space translation symmetry is broken. Quantum mechanics tells us that the center of mass of interacting atoms behaves like a free massive particle and in the ground state it must be totally delocalized. The translationally invariant Hamiltonian leads to quantum eigenstates that are also 
transitionally invariant. To {observe a crystalline structure in the single particle probability density,}  this translational symmetry must be broken -- in other words we do not know 
where a crystal is unless we perform a measurement or a system is perturbed by an infinitesimally weak external perturbation. Let us repeat, this process occurs spontaneously and in  {the} real world one deals with {the symmetry broken state already.}  {After the symmetry is broken} the quantum effects do not give up.  However, they are extremely slow and even if a space crystal were isolated it would take thousand of years to see quantum blurring of the crystalline structure. {It should be noted that if one was able to measure relative distances between particles (not particles positions themselves) in a solid state system prepared in the ground state, signatures of regular arrangement of atoms would be observed and such a measurement would not break the continuous space translational symmetry.}

Spontaneous symmetry breaking is a general property of Nature that surrounds us and it is responsible for a wide class of phenomena like magnetization of ferromagnetic materials (rotational symmetry breaking) or Higgs mechanism (breaking of gauge symmetries) \cite{Strocchi2005}. It is related to a situation where equations describing a system possess a symmetry but a system chooses spontaneously a solution that breaks this symmetry. The effect can be identified with a vulnerability of exact symmetric eigenstates to any  infinitesimally weak perturbation.

In 2012 Frank Wilczek proposed an idea of time crystals \cite{Wilczek2012}. He posed a question whether there exist systems which break spontaneously time translation symmetry. In other words if it is possible that a many-body system self-organizes in time and starts spontaneously  to undergo a periodic motion. This kind of self-organization is a quantum effect and should not be mixed with classical self-organization processes that occur in many systems in nature ranging from flashing Asian fireflies to synchronized clapping of an audience that expresses appreciation for an impressive performance \cite{Glass1988,Neda2000}. Wilczek considered a time-independent system and suggested that it can spontaneously turn to periodic motion even in the lowest energy state. It is already known that the original Wilczek idea cannot be realized \cite{Bruno2013b,Watanabe2015}. However, it became an inspiration to other physicists and a novel research field has been opened. 

Exploration of this time {\it territory} started immediately and soon a new version of time crystals was proposed, that is, a discrete or Floquet time crystals \cite{Sacha2015}. That time crystals are related to quantum self-organization of motion of a many-body system that is periodically driven by an external force. Soon other propositions of Floquet time crystals emerged in driven spin systems \cite{Khemani16,ElseFTC}. While one would expect that in a stationary state a system should follow an external driving it turns out that due to the mutual interaction between particles, the system prefers to move on its own and spontaneously switches to a periodic motion with a different period than a period of driving. This kind of time crystallization has been recently realized in laboratories \cite{Zhang2017,Choi2017}. 

In condensed matter physics properties of space crystals are often analyzed with  the help of space periodic potentials --- it is assumed that a space crystal is already formed, i.e. a crystalline structure has emerged. Systems that are periodically driven in time by an external force not only can reveal spontaneous breaking of a discrete time translation symmetry but can also show solid state phenomena in the time domain. This research has been already initiated and shows that Anderson localization or Mott insulator phase in the time domain can be observed \cite{Sacha15a}.

In this review article we describe the present status of the time crystal research. We start with an introduction to the phenomenon of spontaneous space translation symmetry breaking in solid state systems. Then, we switch to the consideration of possibility of spontaneous time translation symmetry breaking and the idea of time crystal. We will report on the on-going research in time-independent systems and in systems that are periodically driven and are able to reveal discrete time crystal phenomena. Next, the research on a possible realization of condensed matter phenomena in the time domain will be described. 

\section{Time crystals: original idea and perspectives}
We introduce here a concept of time crystals which is related to spontaneous breaking of time translation symmetry in analogy to spontaneous space translation symmetry breaking in the formation of space crystals. We begin with the description of space crystals that allows us to explain phenomena that can also be observed in the time domain when a many-body system switches spontaneously to periodic motion realizing the time crystal.
\subsection{Origin of space crystals: spontaneous breaking of space translation symmetry}
Formation of space crystals relies on periodic self-organization of atoms due to their mutual interactions. Under certain conditions  atoms arrange themselves in a periodic lattice that manifests itself in a periodic behavior in space of a probability density for a measurement of a single particle (an electron or an ion). Strictly speaking such a state cannot be the ground state of a many-body atomic system because it breaks the translation symmetry (assuming a typical nondegenerate ground state). To see this, let us consider a solid state Hamiltonian,
\be
H=\sum_{i=1}^N \frac{\vect p_i^2}{2m_i}+\frac12\sum_{i\ne j}^NU_{ij}(\vect r_i-\vect r_j)
\label{solid_state_h},
\ee
that describes $N$ interacting particles in a finite volume $V$ with periodic boundary conditions. If we shift all positions $\vect r_i$ by the same vector $\vect R$, the Hamiltonian does not change because it depends on relative distances between particles only. It means that the system possesses continuous (i.e. vector $\vect R$ can be arbitrary) space translation symmetry; the corresponding (unitary)  space translation  operator ${\cal T}_{\vect R}$ commutes with $H$ and eigenstates $\psi_n(\vect r_1,\dots,\vect r_N)$ of $H$ are also eigenstates of ${\cal T}_{\vect R}$, 
\bea
{\cal T}_{\vect R}\psi_n(\vect r_1,\dots,\vect r_N)
&=&e^{i\varphi}\psi_n(\vect r_1,\dots,\vect r_N).
\label{ta_eigenval}
\eea
Taking into account Eq.~\eqref{ta_eigenval} and the fact that
\bea
{\cal T}_{\vect R}\psi_n(\vect r_1,\dots,\vect r_N)&=&\psi_n(\vect r_1+\vect R,\dots,\vect r_N+\vect R),
\eea
it is easy to show that the probability density for the measurement of a single particle must be uniform in space if a system is prepared in the ground state (or any other nondegenerate eigenstate),
\bea
\rho(\vect r_1+\vect R)&=&\int
d(\vect r_2+\vect R)\dots d(\vect r_N+\vect R) \cr 
&& \times
\left|\psi_n(\vect r_1+\vect R,\dots,\vect r_N+\vect R)\right|^2
\cr =
\int
&d\vect r_2&\dots d\vect r_N\left|\psi_n(\vect r_1,\dots,\vect r_N)\right|^2
= \rho(\vect r_1),
\eea
and no discrete structure is visible. However, crystalline properties can be observed in the two-point correlation function, 
\be
\rho_2(\vect r_1,\vect r_2)=\int
d\vect r_3\dots d\vect r_N\left|\psi_n(\vect r_1,\vect r_2,\vect r_3\dots,\vect r_N)\right|^2,
\ee
because the symmetry does not forbid $\rho_2(\vect r_1,\vect r_2)$ to be non-uniform in space. If, for a fixed $\vect r_1$, the two-point correlation function reveals periodic behavior as a function of $\vect r_2$, then spontaneous breaking of the continuous space translation symmetry to a discrete translation symmetry can be predicted. The correlation function $\rho_2(\vect r_1,\vect r_2)$ tells us what is the probability density for a measurement of the next particle provided the first particle has been detected at $\vect r_1$. Thus, it is enough to measure a single particle in order to see if a crystalline structure emerges. The measurement can be intentional, i.e. performed by apparatus in a laboratory, or simply due to the coupling of a system to its environment and the resulting possible particle losses. 

The breaking  of the space translation symmetry is related to the localization of the centre of mass of a system \cite{Wezel2007}. In the centre of mass coordinate frame it is apparent that the centre of mass degree of freedom decouples from the relative positions' degrees of freedom, i.e. the Hamiltonian (\ref{solid_state_h}) can be written as
\be
H=\frac{\vect P^2}{2mN}+{\rm relative\; degrees \; of\;freedom.}
\ee
The ground state of a system corresponds to the total momentum $\vect P=0$ and is entirely delocalized in the configuration space. Measurement of particle positions leads to the localization of the system and to an emergence of crystalline structures. {It should be stressed that if one performed the measurement not of individual positions of particles but rather the relative distances between them, such a measurement would also reveal crystalline properties of the system without breaking the continuous space translation symmetry.}

When the continuous space translation symmetry is broken to a discrete symmetry one can ask about the lifetime of the symmetry broken state because it is no longer the system eigenstate. In the thermodynamic limit defined as $N\rightarrow\infty$, the volume $V\rightarrow\infty$ but the particle density $N/V={\rm const.}$, the energy of the symmetry broken state is infinitely close to the ground state energy and its lifetime exceeds easily thousands of years. It can be estimated assuming that the center of mass is described by a wave-packet localized on a length scale $\sigma\approx10^{-11}$m, then, the corresponding kinetic energy is $E_k=\frac{h^2}{2mN\sigma^2}$. If such a system is {\it isolated}, the quantum spreading of the wave-packet leads to a delocalization  of a space crystal. However, in order see that the  crystalline structure is blurred, the center of mass must be delocalized on a length scale of the order of the lattice constant of a crystal, i.e. $a\approx 10^{-10}$m, that 
takes 
time of the order of 
$t=a/(h/mN\sigma)\approx 5\cdot 10^4$ years for $mN=1$kg. Fortunately, jewelery of our grandmothers is not isolated from the outside world and ``Diamonds are forever'' (due to the measurement process).

Kinetic energy $E_k$ is extremely close to the ground state energy and about $10^{-26}$ smaller than energy of an optical photon in our example. It implies that breaking of the symmetry practically costs no energy and can be induced by an infinitesimally small perturbation and, therefore, it occurs spontaneously. 

An alternative method to predict the spontaneous breaking of a space translation symmetry is to apply a symmetry breaking perturbation, calculate the ground state of a system for finite $N$, take the thermodynamic limit and finally turn off the perturbation which leads to the symmetry broken state \cite{Anderson1997,Kaplan1989,Koma1993}. 
\subsection{Origin of time crystals: spontaneous breaking of continuous time translation symmetry}
If a time-independent many-body system is prepared in an eigenstate $\psi_n$ corresponding to energy eigenvalue $E_n$, the probability density for detection of particles at a fixed position in the configuration space obviously does not change in time. It is a direct consequence of continuous time translation symmetry of a system. Indeed, time-independent Hamiltonian $H$ commutes with the time translation operator which is simply the evolution operator ${\cal T}_t=e^{-iHt}$ (we assume $\hbar=1$) and eigenstates of $H$ are also eigenstates of ${\cal T}_t$, thus, 
\be
|\psi_n(t)|^2=|{\cal T}_t\psi_n(0)|^2=|e^{-iE_nt}\psi_n(0)|^2=|\psi_n(0)|^2.
\ee

In 2012 Frank Wilczek proposed \cite{Wilczek2012}  that
 time translation symmetry can be spontaneously broken in an analogue way to space translation symmetry breaking in the formation of space crystals. He coined the term time crystal for that phenomenon. If it exists then a time-independent many-body system, prepared in the ground state, can switch to a periodic motion in time under an infinitesimally weak perturbation. Experimentally it could be observed as a periodic behaviour in time of the probability density for a measurement of a system at a fixed point of the configuration space. It means that switching from space to time crystals we have to exchange the role of space and time. In the space crystal case we expect periodic behaviour in space at a fixed instant of time (i.e. at the moment when we perform a measurement of a system) while in the time crystal case we fix the position in space and ask whether a detector clicks periodically in time. More formal definitions of time crystals may be formulated, see e.g. \cite{ElseFTC,Khemani2017} -- we shall 
stick to this intuitive one.

The idea of time crystals was proposed in two variants. Alfred Shapere and Frank Wilczek \cite{Shapere2012} showed that a classical system can reveal periodic motion in the lowest energy state, {see also \cite{Ghosh2014},} while Wilczek himself \cite{Wilczek2012} presented an idea of quantum time crystal. In the classical case if we ask whether a time-independent system can move and at the same time possess the lowest energy, the answer seems to be obviously No! Indeed, in order to find the lowest energy for a classical particle we have to find an extremal value of a Hamiltonian,
\bea
\frac{\partial H}{\partial p}=0, \quad \frac{\partial H}{\partial x}=0,
\label{classical_c}
\eea
but that  means no motion of a particle is possible because the first condition in (\ref{classical_c}) implies that the Hamilton equation,
\be
\dot x=\frac{dx}{dt}=\frac{\partial H}{\partial p}=0. 
\label{hamilton_e}
\ee
However, if we assume the energy of a particle of the form
\be
E=\frac{\dot x^4}{4}-\frac{\dot x^2}{2},
\label{energy_c}
\ee
one sees that the lowest energy corresponds to particle motion with velocity $\dot x=\pm 1$. This apparent contradiction with the conclusion based on the Hamilton equation (\ref{hamilton_e}) can be resolved when we realize that the energy (\ref{energy_c}) cannot be converted to the Hamiltonian smoothly. That is, the Hamiltonian is a multi-valued function of the momentum with cusps corresponding precisely to energy minima at $\dot x=\pm 1$ where the Hamilton equations are not defined \cite{Shapere2012}. 

Encouraged by the classical analysis we can now look for a quantum time-independent many-body system that in the ground state can spontaneously switch to periodic motion and reveal crystalline properties in the time domain.  Shapere and Wilczek \cite{Shapere2012a} proposed how to quantize the classical single-particle system described by energy (\ref{energy_c}). However, in the following we will concentrate on a more general problem of  many-body systems with more conventional kinetic energy terms that could reveal periodic motion in measurements repeated many times on the same realization of a system. That is, we would like to consider a situation where a detector placed at a certain point in the configuration space reacts periodically because particles are returning to this point. 

A potential example fulfilling our requirements seems to be a superconducting device where an external magnetic field induces current of Cooper pairs. However, the flow of Cooper pairs is uniform and if, at a certain position in space, we detect particles at a certain moment of time, the next detection event will not be correlated temporally with the previous one and no periodic crystalline structure in time will be observed \cite{Wilczek2012,Yamamoto2015}. The original idea of Wilczek was more involved. It is known that interacting particles can form spontaneously inhomogeneous structures in space. On the other hand a charged particle on a ring [one-dimensional (1D) problem with periodic boundary conditions] subjected to magnetic flux can reveal non-vanishing probability current along the ring in the ground state if the flux is properly chosen. Combining these two observations, it should be possible to observe a spontaneous process where a many-body system prepared in the ground state switches to periodic 
evolution where 
an inhomogeneous particles density moves around a ring. Wilczek concentrated on bosons interacting via attractive contact potential on an Aharonov-Bohm ring and we will elaborate on this system in a moment. Very soon another proposition \cite{Li2012}  was suggested that ions on a ring may spontaneously form a space crystal {when kinetic energy of ions is much smaller than Coulomb potential energy between them. Such a Wigner crystal,}  in the presence of a magnetic flux, reveals periodic motion along the ring even if the system is initially prepared in the ground state. Both proposals were immediately criticized by Patrick Bruno \cite{Bruno2013,Bruno2013a} who pointed out that such scenario is impossible --- for replies see \cite{Wilczek2013a,Li2012a}. Soon after Patrick Bruno showed that under quite general conditions spontaneous breaking of continuous time translation symmetry is not possible in any time-independent system prepared in the ground state \cite{Bruno2013b}. Before we present the arguments of Bruno and 
other researchers we first describe details of Wilczek idea \cite{Wilczek2012}. 

Let us consider $N$ bosons on a ring of unit length with attractive contact Dirac $\delta$-interactions and in the presence of a {\it magnetic} flux $\alpha$ (a ring problem in the presence of a magnetic flux is dubbed Aharonov-Bohm ring). The particle mass and $\hbar$ are assumed to be equal to unity and the parameter $g_0$ that determines the strength of attractive interactions is negative. The Hamiltonian of the system,
\be
H=\sum_{i=1}^N\frac{(p_i-\alpha)^2}{2}+\frac{g_0}{2}\sum_{i\ne j}\delta(x_i-x_j),
\label{h}
\ee
possesses continuous time and space translation symmetries and consequently probability density corresponding to any eigenstate is invariant under any translation in time and any translation of all particles along the ring. 

Let us assume for a moment that the {\it magnetic} flux $\alpha=0$ and let us apply the mean field approximation. In the mean field approach all bosons are supposed to occupy the same single particle state $\phi$, i.e., form a Bose-Einstein condensate. Many-body eigenstates are then of the form of a product state $\phi(x_1)\phi(x_2)\dots\phi(x_N)$. In order to obtain the ground state within the mean field approximation one has to find the minimal  value of the energy of the system within the Hilbert subspace spanned by product states that reduces to the solution of the Gross-Pitaevskii equation \cite{Pethick2002},
 \be
\left(-\frac12  \partial_x^2+g_0(N-1)|\phi|^2\right)\phi=\mu\phi.
\label{gpe}
\ee
If the attractive particle interactions are sufficiently strong, i.e. $g_0(N-1)<-\pi^2$, it is energetically favorable to group particles together and the mean field solution $\phi$ breaks the space translation symmetry and becomes inhomogeneous in space \cite{Carr2000}. The $\phi$ function is given by the Jacobi elliptic function but when $|g_0|N\gg 1$, it is well approximated by a bright soliton solution
\be
\phi(x)\approx\cosh^{-1}\left[\frac{g_0(N-1)}{2}(x-x_{\rm CM})\right],
\label{bright_s}
\ee
where $x_{\rm CM}$ is a parameter that describes the centre of mass position \cite{Carr2000,Pethick2002}. Thus, the mean field approach predicts that bosons form a Bose-Einstein condensate where all particles occupy a localized wavefunction (\ref{bright_s}). When we return to the many-body description we obtain that the many-body ground state has the space translation symmetry but this state is strongly vulnerable to any perturbation. In fact to break the space translation invariance it is enough to measure a position of a single particle \cite{Delande2013}. 

Wilczek expected that in the presence of a properly chosen {\it magnetic} flux $\alpha$, one would not only observe spontaneous localization of density of particles in the configuration space but such a density will also move periodically along the ring and the motion sustains forever in the limit when $N\rightarrow\infty$, $g_0\rightarrow 0$ but $g_0(N-1)={\rm const.}$, i.e. in the limit where the mean field prediction remains unchanged, cf. (\ref{bright_s}). This is actually false as  immediately pointed out by Bruno \cite{Bruno2013}. Probably, the simplest way to demonstrate it, is to switch to the centre of mass coordinate frame \cite{Syrwid2017}. Then, the Hamiltonian (\ref{h}) reads
\be
H=\frac{(P-N\alpha)^2}{2N}+{\rm relative\; degrees \; of\;freedom},
\label{h_cm}
\ee
where $P$ is the centre of mass momentum, i.e. the total momentum of the system, which is a conserved quantity. The centre of mass and the relative positions' degrees of freedom decouple and eigenstates of the system are determined by an independent choice of the centre of mass momentum $P_j=2\pi j$ (where $j$ is integer) and the relative degrees of freedom quantum numbers. The ground state corresponds to 
\be
\frac{\partial H}{\partial P_j}=2\pi\frac{j}{N}-\alpha\approx 0.
\label{zero_c}
\ee
In the limit when $N\rightarrow\infty$, Eq.~(\ref{zero_c}) can be fulfilled exactly. This is very bad news because it means there is no probability current related to the centre of mass degree of freedom if the system is prepared in the ground state. Wilczek idea relied on the assumption that if the flux $\alpha$ is chosen properly, quantized values of particle momenta do not allow  Eq.~(\ref{zero_c}) to vanish and once the bright soliton is formed in the process of spontaneous breaking of space translation symmetry it will move too. We see it is not the case in the crucial limit, i.e., where the total number of particles increases.
The idea of Li and co-workers was similar \cite{Li2012}. Instead of particles interacting via an attractive contact potential, they envision ions on an Aharanov-Bohm ring. A spontaneous breaking of space translation symmetry results in a formation of a space crystal which has been supposed to move in the presence of a magnetic flux even if an experiment started with the ground state. Again, the same line of arguments leads to the conclusion that it is not possible.

The impossibility of realization of the time crystal idea was also considered by Nozier\`es \cite{Nozieres13} who investigated a superfluid ring in a magnetic field and presented arguments that a charge density wave cannot reveal rotation induced by diamagnetic currents. However, general analysis of the impossibility of spontaneous time translation symmetry breaking in the ground state was performed by Bruno \cite{Bruno2013b} and later by Watanabe and Oshikawa \cite{Watanabe2015}. Bruno considered a general many-body system on an Aharonov-Bohm ring that is subjected to a perturbation that rotates periodically along the ring. He showed that for the ground state that breaks rotational symmetry, the moment of inertia of the system is always positive and imposing rotation increases energy. Bruno also considered the thermal equilibrium state in the rotating frame, where the Hamiltonian becomes time-independent, and reached a similar conclusion. However, for a periodically driven system analysis of thermal 
equilibrium in a 
rotating frame should be performed with caution because it actually implies that a system is assumed to be in contact with a reservoir of rotating particles. Moreover, it was shown that spontaneous emission of photons may correspond to a jump of an electron in an atom upwards in energy if the process is described in the rotating frame \cite{Delande1998}. This obstacle was overcome by Watanabe and Oshikawa \cite{Watanabe2015} who did not assume the presence of a symmetry broken perturbation but focused on analysis of a correlation function. They proved that the two-point correlation function does not reveal any time dependence when a many-body system is prepared in the ground state or thermal equilibrium state in the limit when volume $V$ of a system goes to infinity.

In the case of space crystals, spontaneous space translation symmetry breaking is demonstrated in the limit when a number of particles $N$ and $V$ tend to infinity but the density of particles is constant. Then, any perturbation is sufficient to break the symmetry and a symmetry broken state lives forever. In the case of time crystals, it is not necessary to take $V\rightarrow\infty$ because we do not have to assume periodic or any other behaviour of a system in space. We may set $V={\rm const.}$, increase $N$ but keep the product of the coupling constant $g_0$ and $N$ fixed, cf. Eq.~(\ref{h}). 

Evolution of the phase of a Bose-Einstein condensate described within ground canonical ensemble can be considered in the context of time crystals. In the ground canonical formalism, the order parameter of a Bose-Einstein condensate reveals periodic oscillations $\la\hat\psi(\vect r,t)\ra=\psi_0e^{-i\mu t}$, where $\hat \psi$ is a bosonic field operator and $\mu$ is a chemical potential \cite{Pethick2002,Castin1998}. Such a periodic time evolution can be measured provided the condensate is coupled to another condensate. If a system is strictly isolated, i.e. when a number of particles is conserved, there is no reference frame and no time dependence can be detected. Volovik \cite{Volovik2013} analyzed a class of such systems (see also \cite{Wilczek2013,Nicolis2012,Castillo2014,Thies2014}) and introduced two relaxation times: energy relaxation time $\tau_E$ and relaxation time $\tau_N$ related to a total number of particles in our example. If $\tau_N\gg\tau_E$, a system with an average number of particles $N$ 
relatively quickly relaxes to a minimal energy state at fixed $N$ and then slowly relaxes to the equilibrium state where no oscillations are present. Although in the intermediate time $\tau_N\gg t\gg\tau_E$ one can observe breaking of time translation symmetry, a system is not strictly in the equilibrium state. While $\tau_N$ can tend to infinity, the strict limit $\tau_N=\infty$ cannot be taken if one wants to observe the oscillation because then there is no reference frame to measure them.

Similar class of systems is considered by Else, Bauer, and Nayak \cite{Else17prx}, however, the authors assume isolated systems. U(1) symmetries that are spontaneously broken are not exact symmetries but they are related to effective Hamiltonians. Spontaneous breaking of U(1) symmetry results in an order parameter that, according to an effective Hamiltonian, oscillates forever. However, at infinitely long times a system approaches a thermal state determined by a full Hamiltonian where no oscillations are observed. 

The results presented indicate that in a many-body system spontaneous breaking of time translation symmetry cannot be observed if a system is prepared strictly in the equilibrium state. Wilczek idea in its original version turned out to be impossible for realization but it became an inspiration to other researchers. Particular progress has been achieved for systems that break discrete time translation symmetry -- the problem described in Section \ref{discrete}. Before considering it let us consider the possibility to utilize excited states of many-body systems.

\subsection{Excited states as a resource?}

Let us analyze if the spontaneous breaking of time translation symmetry can be observed for a many-body system prepared in an excited eigenstate. In other words we would like to answer the question if a time independent many-body system prepared in an excited eigenstate is able to self-organize in time and switch to periodic motion under infinitesimally weak perturbation. We additionally require that such a situation should not be a theoretical issue only but should be realizable experimentally.   

\begin{figure*}
 \includegraphics[width=1.0\linewidth]{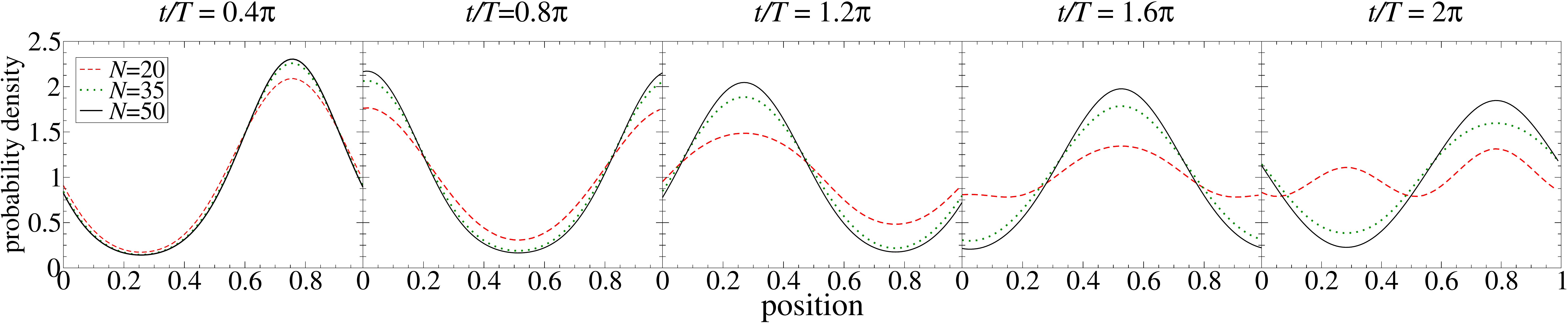}
\caption{ Time evolution of the density-density correlation function \eqref{rho2}, i.e. the probability density for the measurement of the second particle provided the first particle has been measured at $t=0$ at position $x=0.5$. The measurement breaks the continuous translation symmetry making the probability density nonuniform. During a subsequent evolution, as expressed in the different panels, the density moves along a ring with the period $T=1/2\pi$ but also spreads with the characteristic time $t_c$ increasing with the particle number $N$. All results are obtained for $g_0(N-1)=-15$. Reprinted from \cite{Syrwid2017}.} 
\label{fig:evo_sy17}
\end{figure*}

 Analysis of Wilczek model, i.e. the system described by the Hamiltonian (\ref{h}),  \cite{Syrwid2017} suggests that indeed the spontaneous breaking of continues time translation symmetry can occur for an excited eigenstate. For any chosen value of the {\it magnetic} flux $\alpha$, the ground state of the system corresponds to such a total momentum $P_j$ that the probability current associated with the centre of mass motion vanishes if $N\rightarrow\infty$, see Eq.~(\ref{zero_c}). However, if the system is prepared in an eigenstate with total momentum $P_N=2\pi N$, the probability current related to the center of mass motion, 
\be
\frac{\partial H}{\partial P_N}=2\pi-\alpha,
\label{current}
\ee
does not vanish provided the flux $\alpha\ne 2\pi$. Thus, for $\alpha\ne 2\pi$, let us choose among all eigenstates with the total momentum equal $P_N$ the eigenstate $|\psi_0\ra$ with the lowest energy. Although such an eigenstate is not the ground state, it realizes Wilczek's idea. That is, the probability density related to this eigenstate is invariant under time translation transformation but if attractive particle interactions are sufficiently strong, in the limit when $N\rightarrow\infty$, any perturbation can make the density inhomogeneous. The inhomogeneous density should rotate around the ring with the period $T=(2\pi-\alpha)^{-1}$ that is determined by Eq.~(\ref{current}) (remember that the ring has a unit length). 

Numerical simulations \cite{Syrwid2017} have confirmed that the density-density correlation function,
\be\label{rho2}
\rho_2(x,t)\propto\langle\psi_0|\hat\psi^\dagger(x,t)\hat\psi(x,t)\;\hat\psi^\dagger(x_1,0)\hat\psi(x_1,0)|\psi_0\rangle,
\ee
is inhomogeneous as a function of $x$ and reveals periodic rotation around the ring with the lifetime $t_c$ that increases linearly with $N$ if $g_0\rightarrow 0$ but $g_0(N-1)=\rm constant$, i.e., in the limit when the mean field prediction remains unchanged, see Fig.~\ref{fig:evo_sy17}. The density-density correlation function (\ref{rho2}) corresponds actually to the probability density for the measurement of a particle at space point $x$ and at time $t$ provided that at $t=0$ another particle was already detected at $x_1$. It illustrates the nature of a spontaneous process: in order to learn whether the symmetry is broken one has to perform measurement and a minimal possible information, i.e. information about the position of a single particle, is sufficient to break the symmetry. 

Measurement of the position of a single particle results in a localization of the centre of mass of the system on a length scale of the order of the bright soliton size $\sigma_{\rm CM}(0)\approx [g_0(N-1)]^{-1}=\rm const.$, cf. Eq.~(\ref{bright_s}). The localized centre of mass probability density spreads, $\sigma_{\rm CM}(t)\propto t/N$, in the course of time evolution determined by a free particle like Hamiltonian (\ref{h_cm}). Thus, time moment $t_c$ when $\sigma_{\rm CM}(t_c)$ becomes comparable to the length of the ring scales linearly with $N$. That agrees with the numerical results. 
If more particles are measured initially, the centre of mass probability density can be localized on a length scale $\sigma_{\rm CM}(0)\propto N^{-1/2}$. In such a case, the lifetime of periodic evolution of the symmetry broken state scales like $N^{1/2}$ \cite{Syrwid2017}.

It is not simple to prepare in a laboratory a many-body system in a specific excited eigenstate. However, ultra-cold atomic gases constitutes a perfect playground for realization, control and detection not only many-body ground states but also collectively excited states \cite{Pethick2002}. The described spontaneous breaking of continuous time translation symmetry in excited states \cite{Syrwid2017} can be observed in ultra-cold atoms trapped in a toroidal potential in the presence of an artificial gauge field \cite{Goldman2014}. The latter can be realized, e.g., by imposing rotation of a thermal cloud during the evaporative cooling because then the system dissipates to the lowest energy state in the rotating frame and the Coriolis force mimics a magnetic field \cite{Pethick2002}. Proper choice of an artificial gauge potential allows one to obtain the flux $\alpha=2\pi$ for which the ground state is the desire state $|\psi_0\ra$ with the total momentum $P_N=2\pi N$. Then, turning off the gauge 
potential leaves the system in the state $|\psi_0\ra$ which is no longer the ground state of the Hamiltonian with $\alpha=0$ and breaking of time translation symmetry under any weak perturbation is expected. This observation will be in contrast with the same experiment but performed for $g_0(N-1)>-\pi^2$ where no spontaneous breaking of space or time translation symmetry occurs.

To summarize this section, any time-independent many-body system has the continuous time translation symmetry. It implies that probability density corresponding to any eigenstate does not change in time. Moreover, no spontaneous breaking of time translation symmetry can be observed if a system is in the ground state or the thermal equilibrium state. However, it was illustrated with the help of Wilczek's model that the continuous time translation symmetry can be spontaneously broken if a system is prepared in an excited eigenstate. Such a phenomenon could be in principle realized in ultra-cold atoms laboratory. 
The idea of time crystals initiated a novel field of research and inspired many physicists \cite{Chernodub2013,Mendonca2014,Yamamoto2015,Robicheaux2015,Faizal2016}. 
\section{Spontaneous breaking of discrete time translation symmetry: idea and experiments}\label{discrete}
Time independent systems possess continuous time translation symmetry. This symmetry is broken when a Hamiltonian becomes explicitly time dependent $H(t)$. Then, energy is not conserved but if a Hamiltonian is time periodic, $H(t+T)=H(t)$, there exists a kind of stationary states that are time periodic so-called Floquet eigenstates $|u_n(t+T)\ra=|u_n(t)\ra$ \cite{Shirley1965}. Time evolution of any quantum state can be written as a superposition of Floquet states because they form a complete basis at any time,
\be
|\psi(t)\ra=\sum_n c_ne^{-iE_nt}|u_n(t)\ra.
\ee
Substituting $e^{-iE_nt}|u_n(t)\ra$ in the time-dependent Schr\"odinger equation results in an eigenvalue problem for the so-called Floquet Hamiltonian $H_F$, that is hermitian with respect to the scalar product involving integration over time,
\be
H_F|u_n(t)\ra=\left(H(t)-i\partial_t\right)|u_n(t)\ra=E_n|u_n(t)\ra,
\ee
where $|u_n(t)\ra$ must fulfill periodic boundary condition in time. Eigenvalues $E_n$ are real and they are called quasi-energies of a system. These consequences of the Floquet theorem \cite{Shirley1965} are in full analogy to the Bloch theorem known in condensed matter physics where space periodic Hamiltonian are common models of solid state systems. There, momentum of a particle is not conserved because of the presence of an external potential but due to its space periodicity, quasi-momenta are well defined and eigenstates of a particle are plane waves modulated with periodicity of a potential \cite{Ashcroft1976}. In the case of time periodic Hamiltonians we deal with analogues situation, time evolution of a single Floquet state $e^{-iE_nt}|u_n(t)\ra$ nearly agrees with time evolution of an eigenstate of a time-independent system but a time-dependent phase $e^{-iE_nt}$ is modulated with periodicity of a Hamiltonian. Quasi-energy spectrum is not bounded from below. It is actually periodic with a period $2\
pi/T$ and it is sufficient to consider only a single Floquet zone in order to fully describe a system. This is in analogy to a Brillouin zone in condensed matter physics \cite{Ashcroft1976}.

Periodically driven systems break continuous time translation symmetry but they possess discrete time translation symmetry, i.e. a Hamiltonian $H(t+T)=H(t)$ commutes with time translation operator ${\cal T}_T$ related to evolution of a system by period $T$ and Floquet eigenstates are also eigenstates of ${\cal T}_T$, 
\be
{\cal T}_T|u_n(0)\ra=e^{-iE_nT}|u_n(T)\ra=e^{-iE_nT}|u_n(0)\ra.
\ee
Thus, if we choose a point in the configuration space and ask how probability density for detection of a single or many particles at this point changes in time, the answer is it is periodic with a period $T$ if a system is prepared in a Floquet eigenstate. An interesting question arises: can a many-body periodically driven system prepared in a Floquet eigenstate spontaneously self-organize in time and start evolving with a period that is not equal $T$? The answer is yes and this phenomenon, called a discrete or Floquet time crystal, has been recently realized in laboratories \cite{Zhang2017,Choi2017}.

\begin{figure}
 \includegraphics[width=1.0\linewidth]{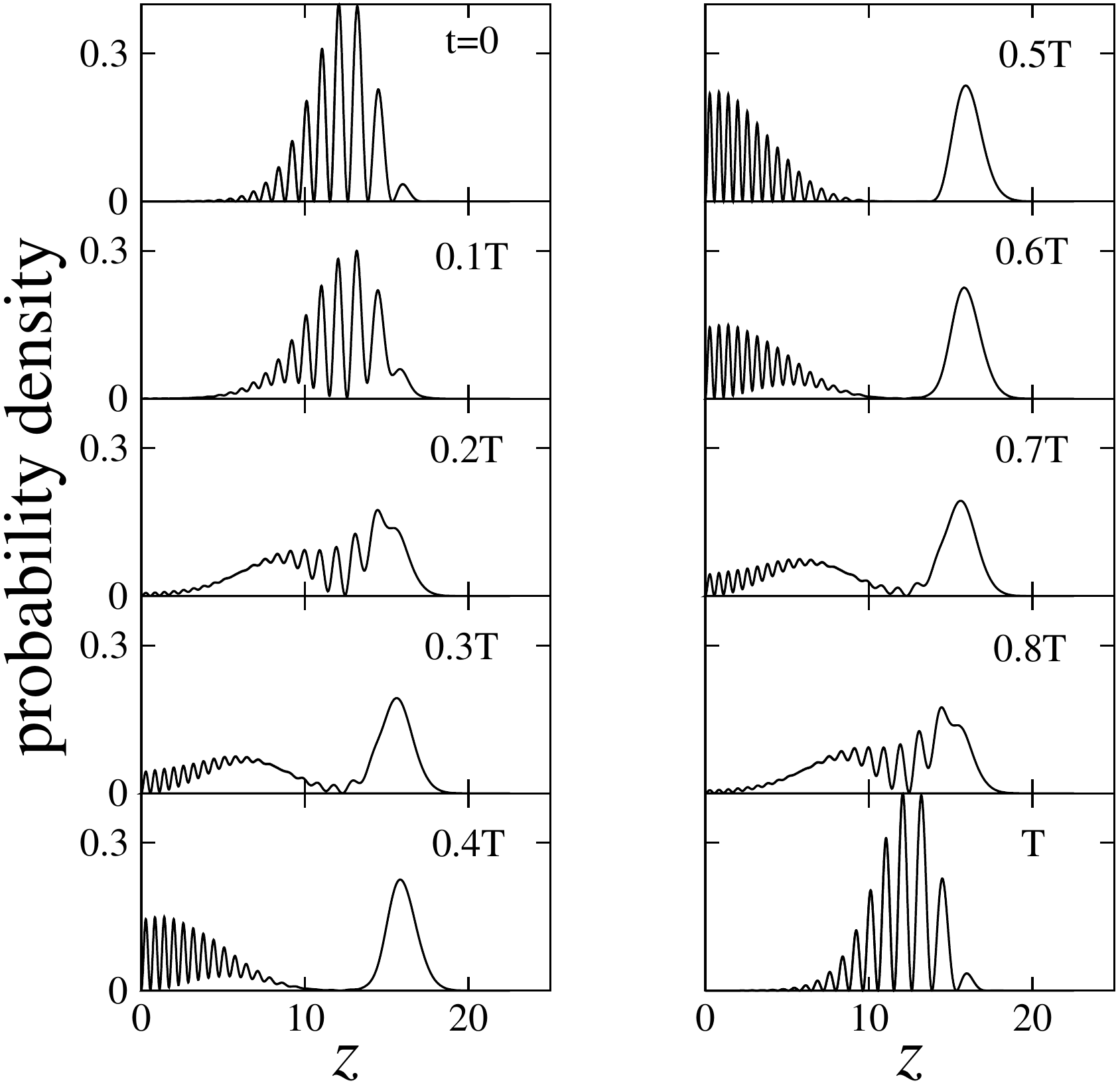}
\caption{Time evolution of the probability density for a particle bouncing on an oscillating mirror and prepared in a Floquet eigenstate that reveals evolution of two localized wave-packet along a classical 2:1 resonant orbit. Different panels correspond to different time moments as indicated in the figure. Initially ($t=0$) two localized wave-packets overlap but because they propagate in opposite directions one can see interference fringes. In the course of the time evolution one wave-packet moves towards the mirror (located at $z=0$) bounces off the mirror and returns. The other wave-packet moves towards the classical turning point in the gravitational field and also returns. Despite the fact {that} each of the wave-packets evolves with a period $2T$, at $t=T$ we end up with the initial situation because at this moment of time the wave-packets exchange their roles.} 
\label{twowave_s15}
\end{figure}

\subsection{Discrete time crystal for atoms bouncing on an oscillating mirror}

The first example of  the spontaneous breaking of discrete time translation symmetry to another discrete symmetry was given in 2015 \cite{Sacha2015}. As an illustration of the effect  a system of ultra-cold atoms bouncing on an oscillating mirror in the presence of the gravitational field is considered. In order to explain the phenomenon we have to first describe the corresponding single particle problem, a classical version of which, called a bouncer has been introduced by Pustyl'nikov as a model for Fermi acceleration \cite{Pustylnikov78,Zaslavsky85}. This famous model of classical chaos has been studied experimentally (see e.g. \cite{Pieranski83}), its quantum version has been often studied \cite{Dembinski93,Buchleitner2002}, moreover, it was realized experimentally for cold atoms bouncing on a mirror formed by an evanescent wave \cite{Steane95}.

A single particle bouncing on an oscillating mirror is described (in 1D approximation and in the frame oscillating with a mirror and in the so-called gravitational units) by the following quantum Hamiltonian \cite{Buchleitner2002},
\be
H=-\frac12 \partial_z^2+V(z)+\lambda z\cos(\omega t),
\label{gpe_s15}
\ee
where $V(z)=z$ for $z\ge 0$ and $V(z)=\infty$ for $z<0$. In the frame oscillating with a mirror, the position of a mirror is fixed (at $z=0$) but the gravitational constant depends periodically on time. In the absence of the mirror oscillations ($\lambda=0$) all classical trajectories of a particle are periodic with a period increasing with the energy of the particle. When mirror oscillations are turned on, classical motion becomes irregular but some of periodic orbits survive. They are stable resonant orbits living in regular parts of the classical phase space. There is 1:1 resonant orbit where a particle moves periodically with a period equal to the period of the mirror oscillations $T=2\pi/\omega$. There exist 2:1 resonant orbit where a particle bounces on a mirror with a period twice longer than that of the oscillations as well as higher order resonances $s:1$. Switching to the quantum description, a motion of a particle is described by Floquet eigenstates. It may be surprising but classical-like motion 
of a particle on resonant orbits can be also observed in the quantum world. For example, 
suitable choice of parameters results in a Floquet eigenstate that is represented by a localized wave-packet moving periodically along 1:1 resonant orbit. This is an example of the so-called non-spreading wave-packet motion that was discovered more than 20 years ago \cite{Henkel1992,Bialynicki1994,Delande1994,Holthaus1995,Buchleitner1995,Zakrzewski1995} for a review see \cite{Buchleitner2002}. 

We will focus on the 2:1 resonance case \cite{Sacha2015}. A single localized wave-packet moving along classical 2:1 resonant orbit cannot form a Floquet eigenstate because it moves with a period twice longer than $T$. However, a superposition of two such wave-packets that move with the period $2T$ but after $T$ exchange their positions can form a proper Floquet eigenstate and indeed there exists such a state, see Fig.~\ref{twowave_s15}. Two localized wave-packets can actually form two mutually orthogonal superpositions. Therefore, there exist two Floquet eigenstates, $u_1(z,t)$ and $u_2(z,t)$, consisting of two localized wave-packets moving along the 2:1 resonant trajectory. The eigenstates $u_1(z,t)$ and $u_2(z,t)$ correspond to nearly degenerate (modulo $\omega/2$) quasi-energies $E_2-E_1=\omega/2+J$. A tiny splitting $J$ is related to a tunneling process. By superposing $u_1(z,t)$ and $e^{-i\omega t/2}u_2(z,t)$ we can eliminate one of the two localized wave-packets. The remaining wave-packet evolves 
along 
2:1 resonant trajectory but at the same time slowly tunnels to a position of the other missing wave-packet --- the full tunneling process is completed after a period $\pi/J$ which is very long as compared to $T$. It is worth noting that in order to observe the described resonant behaviour, the resonant condition does not need  to be strictly fulfilled. That is, if the resonance condition corresponds to an unperturbed classical orbit with a sufficiently high energy, in the quantum description the nearest in energy unperturbed quantum states will form the Floquet eigenstates that we look for. In other words small changes of the driving frequency $\omega$ do not change the behaviour and no fine tuning is necessary because the motion of localized wave-packets is protected by local constants of motion related to regular parts of the classical phase space \cite{Buchleitner2002}.

\begin{figure}
\includegraphics[width=0.5\linewidth]{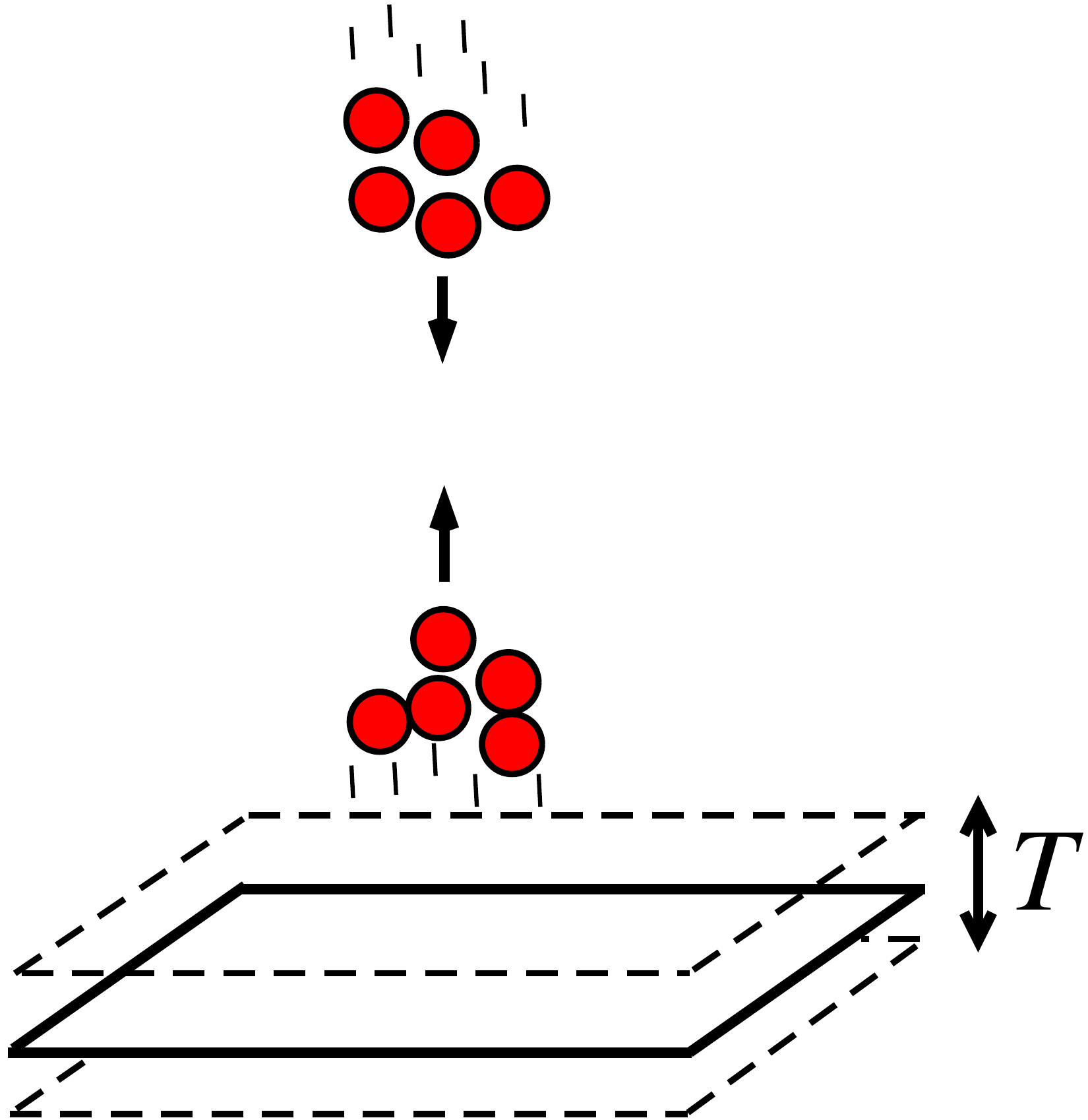}
 \includegraphics[width=0.45\linewidth]{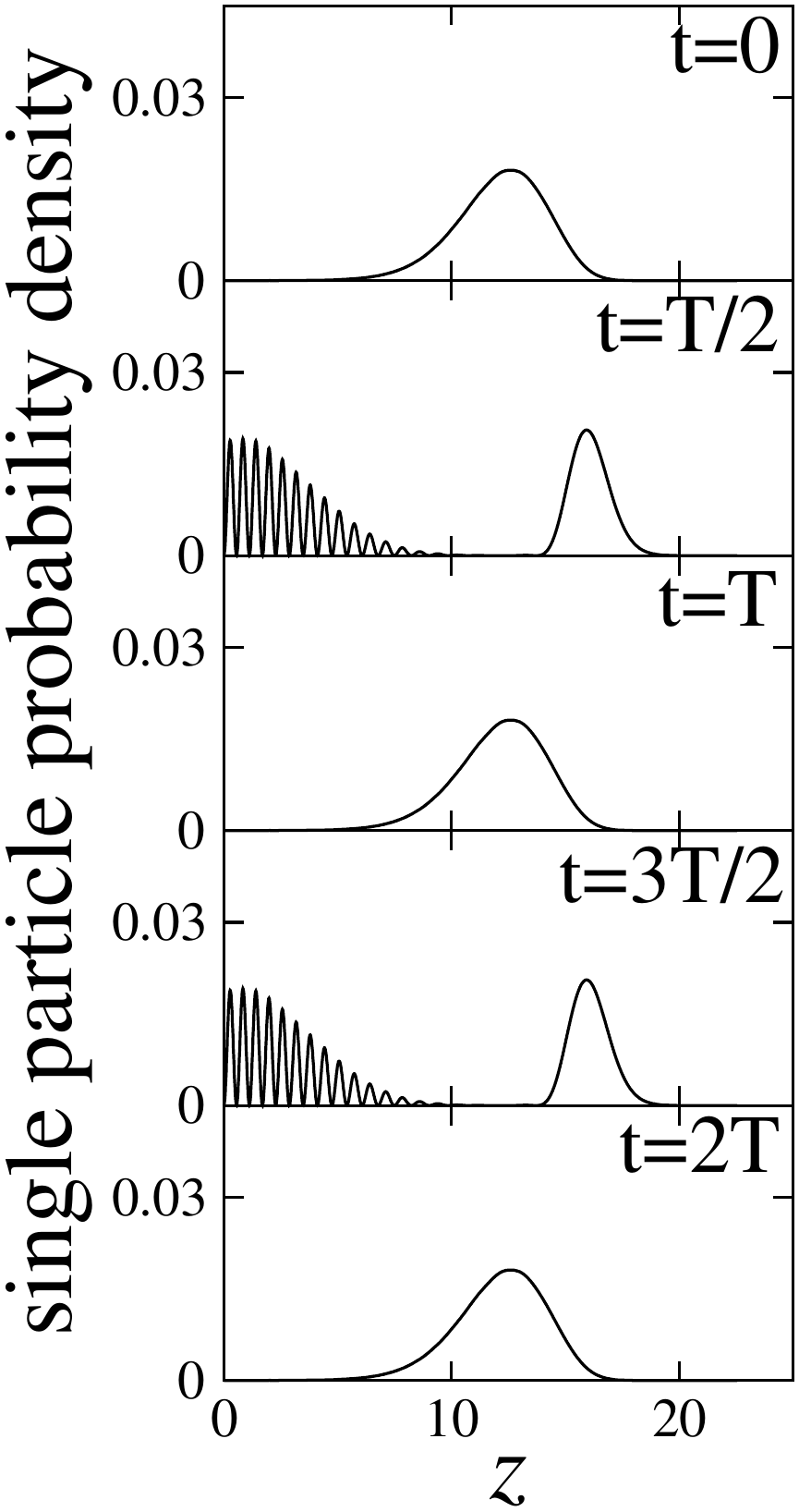}
\caption{Left panel: schematic plot of a system of $N$ atoms bouncing on an oscillating mirror and prepared in a many-body Floquet state (\ref{cat_s15}). Each of two atomic clouds moves with a period $2T$ but they exchange their positions after time $T$ so that the entire Floquet state is periodic with a period $T$. Right panel shows time evolution of the corresponding single particle probability density (\ref{single_p_s15}) for $N=10^4$, $g_0N=-0.5$, $\omega=1.1$ and $\lambda=0.06$. Reprinted {and adapted} from \cite{Sacha2015}.} 
\label{szkic1_s15}
\end{figure}

\begin{figure}
\includegraphics[width=0.5\linewidth]{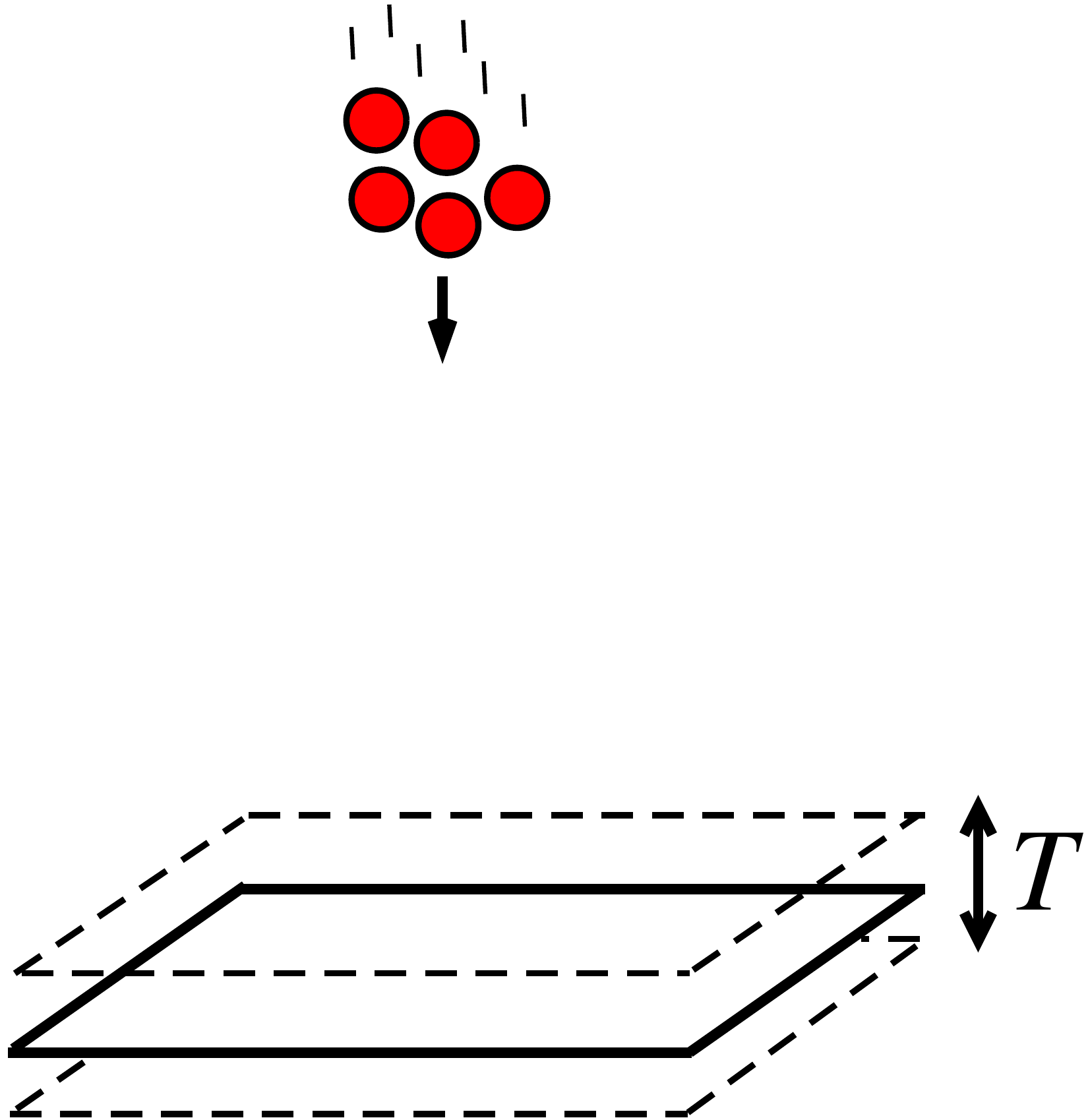}
 \includegraphics[width=0.42\linewidth]{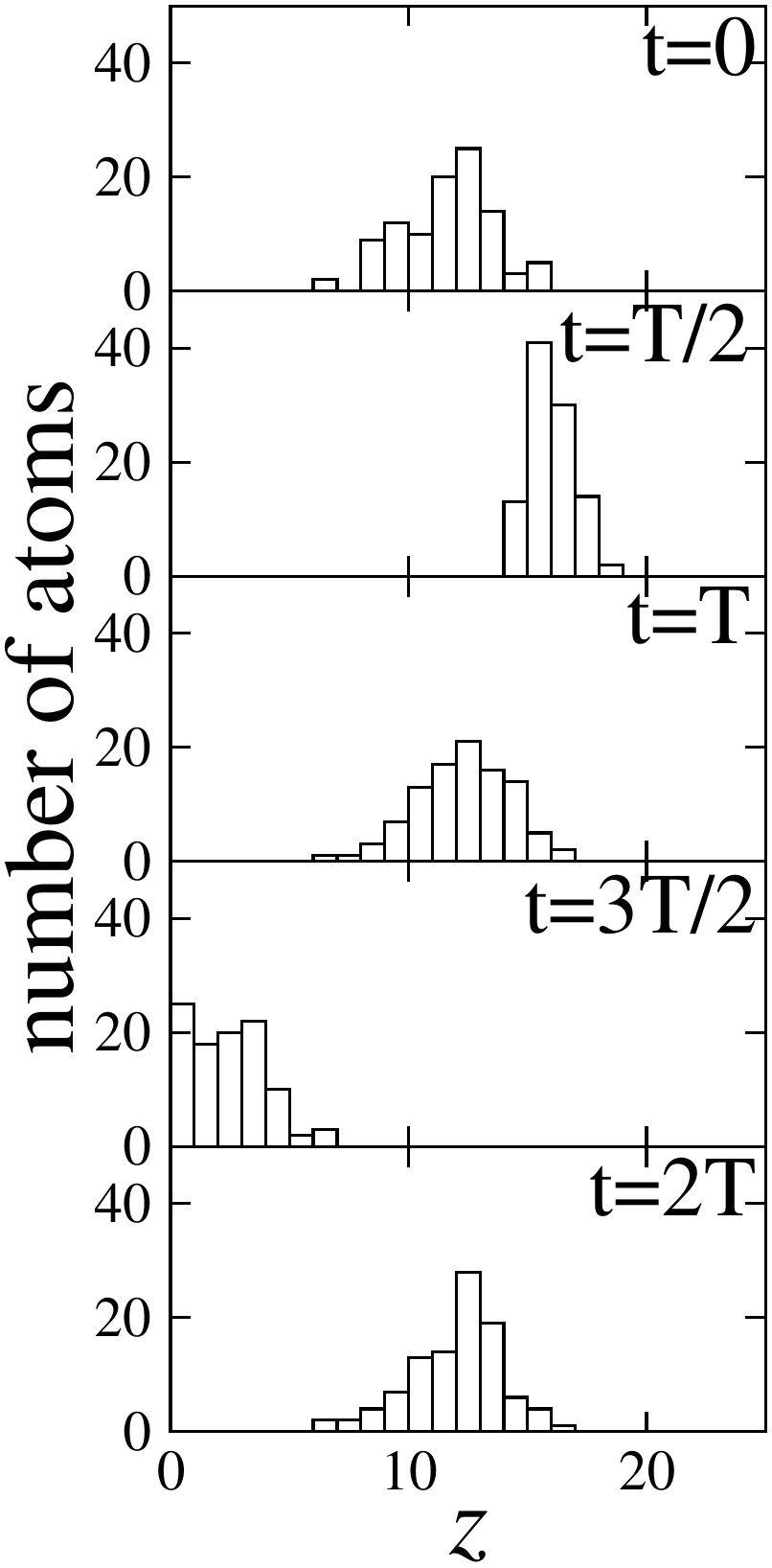}
\caption{Left panel: schematic plot of a system of $N$ atoms bouncing on an oscillating mirror as in Fig.~\ref{szkic1_s15} but after the measurement of the position of a single atom --- atomic cloud visible in the plot moves with a period $2T$. Right panel shows the results of the measurements of positions of 100 atoms, i.e. at $t=0$ one measures positions of 100 atoms, let the remaining atoms evolve and after $T/2$ one again measures positions of 100 atoms and so on. The histograms presented in the right panel indicate that time periodic evolution of the system, after the spontaneous time translation symmetry breaking, can be observed in a single experimental realization. The initial number of atoms $N=10^4$ and the other parameters as in Fig.~\ref{szkic1_s15}. Reprinted {and adapted} from \cite{Sacha2015}.} 
\label{szkic2_s15}
\end{figure}

When we consider $N$ particles bouncing on an oscillating mirror we observe a similar behaviour if particles are bosons and they do not interact. That is, in the Hilbert subspace spanned by Fock states $|n_1,n_2\ra$, where $n_1$ and $n_2=N-n_1$ are numbers of particles occupying single particle Floquet states $u_1$ and $e^{-i\omega t/2}u_2$, respectively, the lowest quasi-energy eigenstate corresponds to $|N,0\ra$. In order to describe particles interacting via $\delta$-contact potential one has to consider the following many-body Floquet Hamiltonian
\bea
H_F&=&\int\limits_0^{2T}dt\int\limits_0^\infty dz\;\hat\psi\left[H(t)+\frac{g_0}{2}\hat\psi^\dagger\hat\psi-i\partial_t\right]\hat\psi,
\cr
&\approx& -\frac{J}{2}\left(\hat c_1^\dagger \hat c_1-\hat c_2^\dagger \hat c_2\right)+\frac14 (U-2U_{12})\cr
&&\times \left[(\hat c_1^\dagger)^2\hat c_2^2+(\hat c_2^\dagger)^2\hat c_1^2+2\hat c_1^\dagger \hat c_1\hat c_2^\dagger\hat c_2\right]+{\rm const.}, \cr &&
\label{manyH_s15}
\eea
where $U$ and $U_{12}$ are equal to integrals over time and space of products of the probability densities of two evolving localized wave-packets multiplied by $g_0$. In Eq.~(\ref{manyH_s15}) we restricted to the Hilbert subspace spanned by the previously described Fock states $|n_1,n_2\ra$ and the bosonic field operator $\hat\psi(z,t)\approx u_1\hat c_1+e^{-i\omega t/2}u_2\hat c_2$ where $\hat c_1$ and $\hat c_2$ are standard annihilation operators.  Such a restriction is valid provided the interaction energy is very small and couplings to the complementary Hilbert subspace can be neglected. It will be the case because the interaction energy per particle will be of the order of the tunneling splitting $J$ that is extremely small as compared to any other energy scale of the system. Eigenstates of the Hamiltonian (\ref{manyH_s15}) correspond to many-body Floquet states that are also eigenstates of the time translation operator ${\cal T}_T$ what is apparent when one realizes that there are two classes of 
the eigenstates of (\ref{manyH_s15}). Eigenstates from the first class are spanned by Fock states with only even occupations of the $e^{-i\omega t/2}u_2(z,t)$ mode while eigenstates from the other class by Fock states with the odd occupations only \cite{Sacha2015}.

Assuming attractive ($g_0<0$) interactions that are very weak, i.e. $N|U-2U_{12}|<J$, the ground state $|\psi_0\ra$ of (\ref{manyH_s15}) matches the non-interacting result, $|\psi_0\ra\approx |N,0\ra$. However, when
\be
N|U-2U_{12}|>J,
\label{cond_s15}
\ee
it is energetically favorable to collect all bosons in a single localized wave-packet and consequently the ground state of the Hamiltonian (\ref{manyH_s15}) is a Schr\"odinger cat-like state which is clear if one writes such a many-body Floquet eigenstate in another Fock basis $|\tilde n_1,\tilde n_2\ra$ where $\tilde n_1$ and $\tilde n_2=N-\tilde n_1$ are occupations of the first localized wave-packet, $\phi_1=(u_1+e^{-i\omega t/2}u_2)/\sqrt{2}$, and the other localized wave-packet, $\phi_2=(u_1-e^{-i\omega t/2}u_2)/\sqrt{2}$, respectively. Then, the many-body ground state reads
\be
|\psi_0\ra\approx\frac{|N,0\ra+|0,N\ra}{\sqrt{2}}.
\label{cat_s15}
\ee
The corresponding single particle probability density,
\bea
\rho_1(z,t)&=&\langle\psi_0|\hat \psi^\dagger(z,t)\hat \psi(z,t)|\psi_0\rangle
\cr &\approx&\frac{N}{2}\left(|\phi_1(z,t)|^2+|\phi_2(z,t)|^2\right),
\label{single_p_s15}
\eea
is plotted in Fig.~\ref{szkic1_s15} at different time moments. 
The discrete time translation symmetry is preserved in time evolution of the many-body Floquet eigenstate $|\psi_0\ra$ but this state is extremely vulnerable to any perturbation. After a measurement of a position $x_1$ of a single particle, the symmetry is gone because the quantum state of the remaining particles immediately collapses to one of the terms in the sum (\ref{cat_s15}), i.e. to $|N-1,0\ra$ or $|0,N-1\ra$ depending on a result $x_1$ of the measurement. Then, time evolution shows that the original discrete time translation symmetry has been broken, i.e. the system evolves with the period $2T$. The resulting state $|N-1,0\ra$ (or $|0,N-1\ra$) is robust against any further  perturbation --- one can perform many measurements and still the period of the time evolution remains $2T$, see Fig.~\ref{szkic2_s15}. Tunneling time from the state $|N-1,0\ra$ to $|0,N-1\ra$ or vice versa, in the limit $N\rightarrow\infty$, $g_0\rightarrow 0$ but $g_0N=\rm const.$, increases like $e^{\alpha N}/N$ with a positive 
constant $\alpha$ 
and very quickly becomes so long that it is not-measurable \cite{Sacha2015}.  

Not only the ground and first excited states of (\ref{manyH_s15}) possess a Schr\"odinger cat-like structure. With $N\rightarrow\infty$ more and more eigenstates come in pairs of cat states \cite{Zin2008,Oles2010} and experimental preparation of any initial state where most of atoms occupy a single localized wave-packet will result in time evolution with the period $2T$ that practically never ends. This is in contrast to the same experiment performed for $N|U-2U_{12}|<J$  where one will observe tunneling of atoms to another localized wave-packet after time period of the order of $1/J$. 

The presented example constitutes an illustration that spontaneous self-organization in time of a periodically driven many-body system is possible, i.e. spontaneous breaking of discrete time translation symmetry to another discrete symmetry may occur. In the following Section we shall discuss a similar behaviour observed for a very different model -- a system of driven interacting spins prepared in such a way that during a single period $T$ all the spins are flipped (being thus brought to the same orientation after $2T$). Then, the corresponding Floquet eigenstates are macroscopic Schr\"odinger cat-like states of two possible orientations of spins. A measurement of the direction of even a single spin leads to a collapse of any of these cat-like states to a short range correlated time crystal state.
\subsection{Discrete time crystals in spin systems}
Interestingly, the original presentation of discrete time translation symmetry breaking \cite{Sacha2015} was not originally noticed. It took more than a year to rediscover that phenomenon in quite a different setting \cite{Khemani16,Keyserlingk16b,ElseFTC,Else17prx} involving, importantly, the disorder induced effects.

Recent years brought an intensive study of many-body localization (MBL) -- a phenomenon which seems to break a common wisdom about disordered many body systems. The latter were supposed to generically thermalize if evolved from some nonstationary initial state. Starting from a seminal work of Basco, Aleiner, and Altschuler \cite{Basko06} it seems it does not have to be the case, a sufficiently strong disorder may bring a many body system to a localized, non-ergodic phase \cite{Oganesyan07,Znidaric08}. The phenomenon received a lot of attention with literally hundreds of publications in last 10 years (for excellent early reviews see \cite{Huse2014,Rahul15}). 

While mainly observed for one-dimensional spin chains, MBL is considered by now to be a generic phenomenon in disordered many-body systems. 
The latter are often studied using strong periodic driving -- being a natural current extension of strongly periodically driven single particle systems intensively studied in the last millennium (consider e.g. laser-atom interactions or microwave resonance techniques). Until quite recently it has been a common understanding that periodic driving of a many body system must supply (on average) energy to the system and lead to 
heating (see e.g. \cite{Alessio14,Lazarides14a,Lazarides14}) for arbitrary initial states -- see, however, \cite{Abanin2015,Chandran16,Mori2016,Kuwahara2016}. Since any periodically driven system may be described by Floquet time-periodic eigenvectors \cite{Shirley1965,Sambe1973}, the inevitable heating suggests a lack of possibility to prepare an initial state in a form of a single, or few Floquet eigenstates. Interestingly, this believe is wrong for simple driven single particle problems where specific localized examples -- the so-called nonspreading wavepackets \cite{Buchleitner2002} were mentioned already in the previous Section. They have been also demonstrated experimentally \cite{Maeda2004}. Let us note that lifetime of states that are superpositions of few Floquet eigenstates is infinite. These wavepackets were studied, however, in relatively 
simple small systems while the 
heating argument was 
usually presented in the 
thermodynamic limit.

Numerical experiments revealed that in the presence of disorder MBL persists for driven systems provided the frequency of the driving is high enough \cite{Ponte2015,Abanin2015,Lazarides15} so the long-time system behavior may be analyzed by an effective Hamiltonian resulting from proper time averaging over the period of the drive. Soon it has been realized that periodically driven systems can not only remain many-body localized but that one can distinguish different ``phases'' by means of appropriate correlators as discussed for spin systems by \cite{Khemani16}. Among those, in the present context, important is the phase named as $\pi$-spin glass.  While the original analysis
\cite{Khemani16} uses (Jordan-Wigner transformation based) link to earlier works on driven  Majorana edge modes for noninteracting systems \cite{Jiang2011,Bastidas2012,Dutta2013} let us stay within the language of Schr\"odinger cat-like delocalized Floquet states as discussed in the previous Section. 

The system considered \cite{Khemani16} is a driven spin model with a binary periodic drive corresponding to the unitary evolution over the period $T=t_1+t_2$ given by $U=\exp(-it_2H_x)exp(-it_1H_z)$ with
\bea \label{hamkhe}
H_z&=&\sum_ih_i\sigma_i^z+\sum_iJ_z\sigma_i^z\sigma_{i+1}^z \nonumber \\
H_x&=&\sum_i \left(J_i\sigma_i^x\sigma_{i+1}^x +J_z\sigma_i^z\sigma_{i+1}^z\right). 
\eea
Aiming at discussing physics in the many-body localization regime, the authors analyse statistics of quasi-energies (eigenvalues of $U$) looking in particular at 
the average $\bar r$ ratio between the smallest and the largest adjacent energy gaps \cite{Oganesyan07}:
$r_n = \min[\delta^\epsilon_n ,\delta^\epsilon_{n-1}]/ \max[\delta^\epsilon_n,\delta^\epsilon_{n-1}],$ with $\delta^\epsilon_n=
\epsilon_n - \epsilon_{n-1},$ and $\epsilon_n\in [0,2\pi)$ are the quasi-energies. 
In the MBL phase, one expects $\bar r$ to be close to the Poisson limit $\bar r^\mathrm{Poisson}=2\ln 2 -1\approx0.386$
\cite{Atas13}. That allows one to choose the proper values for $J_z$ and suitable distributions for random values of $h_it_1$ and $J_it_2$ in \eqref{hamkhe}. 

\begin{figure}
 \includegraphics[width=1.0\linewidth]{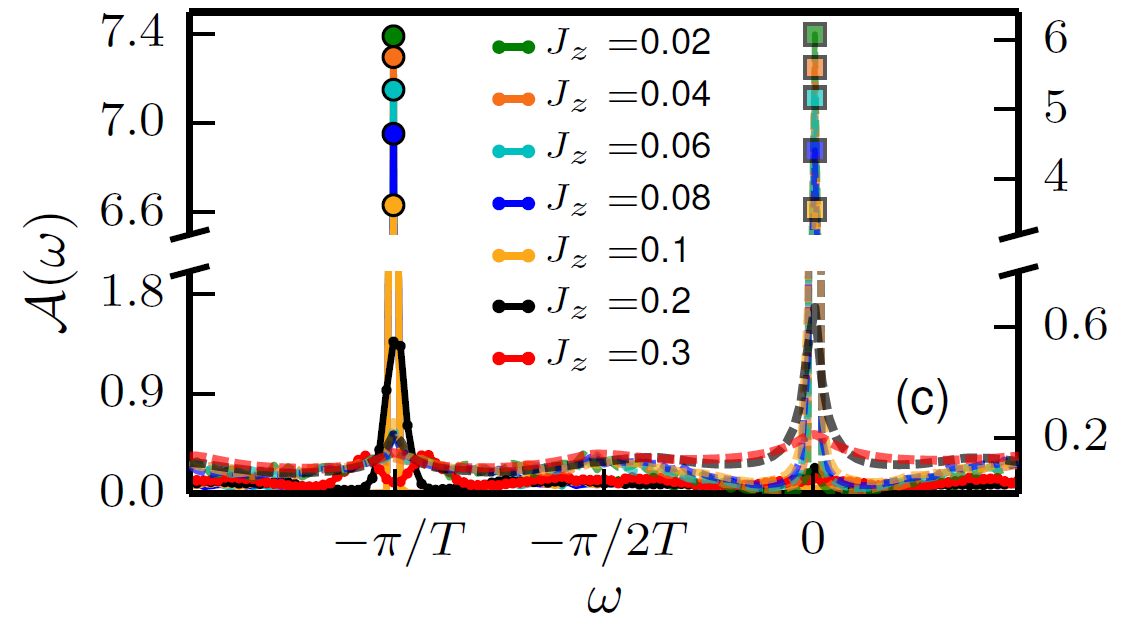}
\caption{Disorder averaged spectral function \eqref{eq:khe} for different values of the interaction term in \eqref{hamkhe}. Solid lines, corresponding to $\pi$-spin glass phase, reveal a robust peak at $-\pi/T$ with $T$ being the period of the drive -- in (\ref{hamkhe}), $h_it_1$ and $J_it_2$ are chosen randomly and uniformly in the intervals $[1.512,1.551]$ and $[0.393,1.492]$, respectively. Reprinted figure with permission from V. Khemani, A. Lazarides, R. Moessner,  and S.L. Sondhi, Phys. Rev. Lett, {\bf 116},250401 (2016). Copyright (2016) by the American Physical Society \cite{Khemani16}.}
\label{khemani2}
\end{figure}

Observe that $H_x$ combined with $H_z$ enjoys similar symmetries to Ising model. The Floquet quasi-energy eigenstates are also eigenstates of parity $P=\Pi_i \sigma^z_i$.
Observe also that for $h_it_1=\pi/2$ the role of the first term in $H_z$ will be to flip the spins in the $x$ direction. Neglecting for a moment the other terms in $U$, its action in two periods of the drive brings the spins to the same orientation. One then expects that many-body localized (when disorder and interactions are included) Floquet eigenstates will resemble Schr\"odinger cat states
with frozen domain walls deep in the spin glass phase {
\be|\pm\ra= \frac{1}{\sqrt{2}}\left(|\uparrow\downarrow\uparrow\dots \uparrow\ra_x\pm|\downarrow\uparrow\downarrow\dots\downarrow\ra_x\right).
\ee
}They will come in pairs with different parity being  almost degenerate in energy (modulo $\pi/T$) 
 revealing the existence of $\pi/T$  phases resulting from discrete symmetry breaking (that destroys vulnerable cats). 
 Fig.~\ref{khemani2} presents the disorder averaged spectral function of the spin raising operator, $\sigma_i^+$ on an arbitrary site $i$ in the Floquet basis
 \be\label{eq:khe}
 {\cal A}(\omega)=\frac{1}{2^L}\sum_{\alpha\beta}\langle\phi_\alpha(0)|\sigma^+_i|\phi_\beta(0)\ra\;\delta(\omega-(\epsilon_\alpha-\epsilon_\beta)),
 \ee 
 where $\epsilon_\alpha$ are Floquet eigenvalues with corresponding vectors $|\phi_\alpha\rangle$ with $L$ being the system size. A peak at $-\pi/T$ is robust against changes of interaction strengths, disappearing only  above a critical interaction strength. The effective period doubling observed may be interpreted as a potential Floquet realization of a time crystal \cite{Khemani16}.

The same group presented a nice general group analysis of symmetry broken phases in Floquet systems \cite{Keyserlingk16b,Keyserlingk16} in a similar setting. This time
the authors chose to flip the role of $x$ and $z$ axes considering the model 
 defined by $U=U_2U_1=\exp(-it_2H_2)\exp(-it_1H_1)$ with the period $T=t_1+t_2$ and
\bea \label{ham2}
H_1&=&\sum_ih_i\sigma_i^x\nonumber \\
H_2&=&\sum_iJ_i\sigma_i^z\sigma_{i+1}^z, 
\eea
where $h_i$ may be random while $J_i$ are uniformly drawn from  $[J_z-\delta J, J_z+\delta J]$ (to obtain many-body localization). This is a ``minimal'' spin system with $J_z$ controlling the interactions. Again for $h_i t_1=\pi/2$  the action of $\exp(-it_1H_1)$ corresponds to flipping the spins along, this time,  $z$ direction, i.e., $U_1=\Pi_ri\sigma^x_r$. The second unitary is diagonal in $z$ components. Clearly the quasi-energy eigenstates will be now cat-like states {
\be | \pm\rangle\propto |\{d_i\}, p=\pm\rangle=  \frac{1}{\sqrt{2}}\left(|\{m_i\}\rangle_z \pm |\{-m_i\}\rangle_z\right),
\label{eqcat}
\ee
}where $d_i$ is the expectation value of $\sigma_i^z\sigma_{i+1}^z$ and $p$ is the Ising parity of the state. While the argument is presented for $h_it_1=\pi/2$ the numerical evidence \cite{Khemani16} reveals that the phase obtained is stable and robust. It is shown that  broken symmetry phases are stable against  weak local deformations of Floquet drives and it is stressed that the order revealed by $\pi$-spin glass (i.e. Floquet time crystal) is always spatio-temporal and never purely temporal as related to general symmetry properties -- for details see  \cite{Keyserlingk16}. The authors stress that absolutely stable states are generally possible by a combination of Floquet periodicity, broken symmetries and MBL -- on the other 
hand 
no proof is given that MBL is a 
necessary condition for observing absolutely stable phases. It is shown that signatures of stable time crystals may be obtained  starting from generic short-range initial states.

In a number of papers \cite{Keyserlingk16a,Else16,Potter2016,Po2016,Roy2016,Harper2017} a further classifications of possible phases in periodically driven system have been attempted with the stress on novel interesting topological opportunities offered by Floquet Hamiltonians. This goes beyond the modest scope of this review centered on time crystals so we only mention those papers for an interested reader.  
 
An independent  clarification of time crystals  for spin systems has been made by Else, Bauer, and Nayak \cite{ElseFTC}. They consider a general time-periodic Hamiltonian and formulate simple criteria for  the occurrence of a discrete time translation symmetry breaking (DTTSB) in terms of Floquet eigenstate properties. A simple statement is that Floquet eigenstates cannot be short-range correlated. Note that, as nicely pointed out \cite{ElseFTC}, this condition implies lack of MBL in Floquet systems exhibiting DTTSB if MBL is strictly understood as an integrable disordered system with a complete set of local integrals of motion (LIOMs) \cite{Huse2014,Serbyn2013}.
 Non local  Floquet states can not be all the common eigenstates of the Hamiltonian and the quasi-local operators so the system of LIOMs (together with the Floquet Hamiltonian) cannot form a complete set of observables. As shown \cite{ElseFTC} the condition of non local character of Floquet eigenstates is equivalent to the statement that for any time instant, $\tau$ and any state $|\psi(\tau)\rangle$ with short-range correlations there exist an operator  $\hat O$ with the average in state $\psi(\tau)$ being not time-periodic with period $T$ despite $T$-periodicity of the Hamiltonian itself. Note that this property is operational, i.e. it suggests that DTTSB can be detected by measurements on such an operator.
 
\begin{figure}
 \includegraphics[width=1.0\linewidth]{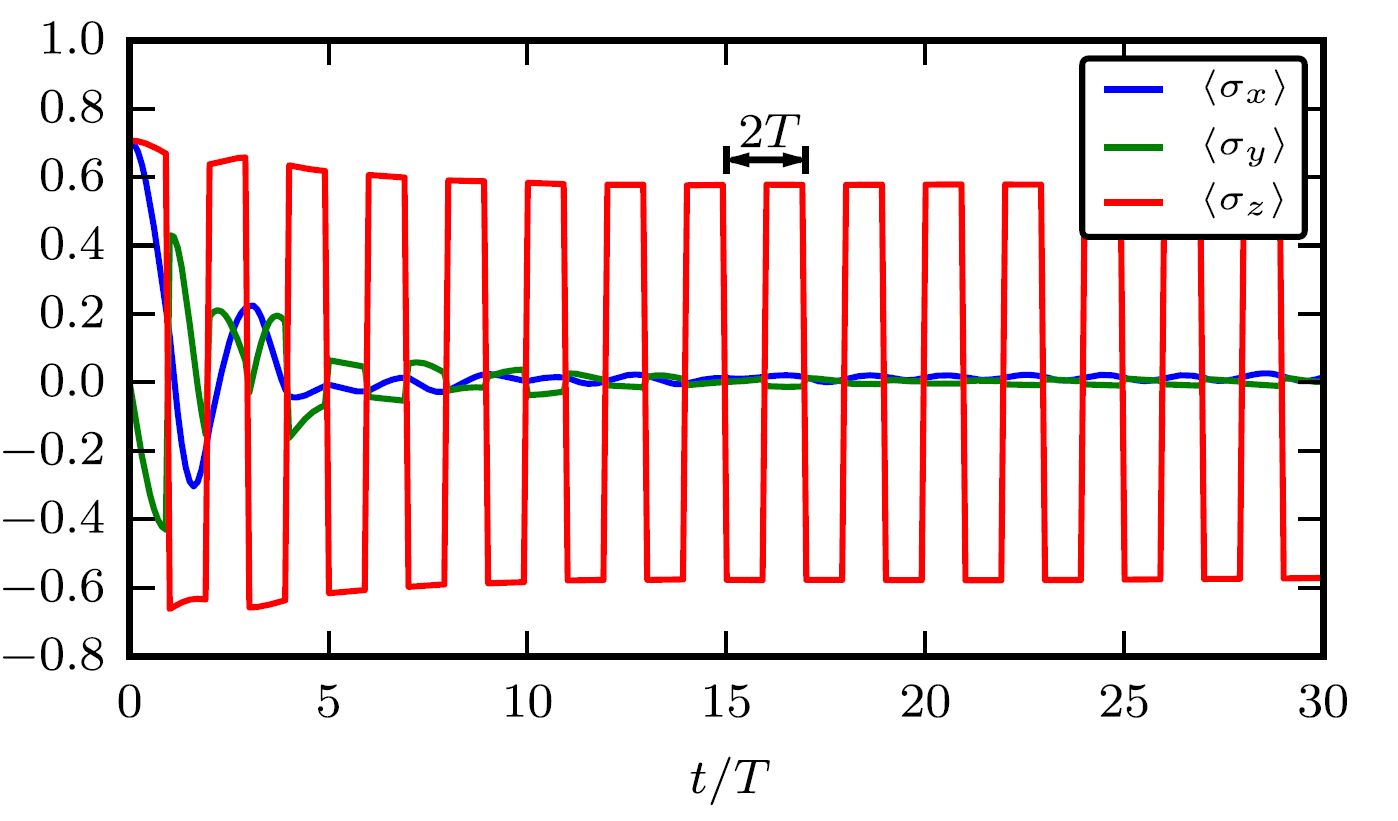}
\caption{The time-dependence of the disorder averaged magnetizations $\langle \sigma_i^\alpha\rangle\equiv \langle\sigma_\alpha\rangle$ as indicated in the figure. Magnetization along $z$ axis shows persistent oscillations with twice the period $T$ of the driving -- compare \eqref{eqelse0}-\eqref{eqelse2}. Exact $\pi$ spin flip is assumed as in \eqref{eqelse0}.   Reprinted figure with permission from D. V. Else, B. Bauer,  and C. Nayak, Phys. Rev. Lett, {\bf 117}, 090402 (2016). Copyright (2016) by the American Physical Society \cite{ElseFTC}} 
\label{nayak1}
\end{figure}

The example studied, a simple spin system (very close to that considered by \cite{Khemani16,Keyserlingk16b,Keyserlingk16}) applies within a single period two operations, a global rotation around $x$ axis by $\pi$ in time  $t_1=\pi/2$ with the unitary 
\be\label{eqelse0}
U_1=\exp\left[it_1\sum_i \sigma_x\right],
\ee
followed by an evolution for time $t_2$ in a disordered Hamiltonian $H_{MBL}$,  
\be \label{eqelse1} U_2=\exp\left[-iH_{MBL}t_2\right], \ee
with $T=t_1+t_2$ and 
\be \label{eqelse2}
H_{MBL}= \sum_i \left(J_i\sigma_i^z\sigma_{i+1}^z+h_i^z\sigma_i^z+h_i^x\sigma_i^x\right).
\ee
The parameters $J_i,\ h_i^z$, and $h_i^x$ are uniformly and randomly chosen, $J_i\in[J/2,3J/2]$, $h_i^z\in[0,h^z]$, and $h_i^x\in[0,h]$ with, typically $J=h^z=1$ and $h\ll J$. At the prize of repeating the reasoning presented above, we mention that the system is  soluble for $h=0$, the eigenstates of $H_{MBL}$ being the product of eigenvectors of individual $\sigma_i^z$, { $|\{m_i\}\rangle_z$,  $\sigma_k^z|\{m_i\}\rangle_z=m_k|\{m_i\}
\rangle_z$} with $m_k=\pm 1$. 
Explicitly {
\be 
H_{MBL}|\{m_i\}\rangle_z = [E_2(\{m_i\}) +E_1(\{m_i\})]\; |\{m_i\}\rangle_z,
\ee		
}with $E_2(\{m_i\})=\sum_i J_im_im_{i+1}$ and  $E_1(\{m_i\})=\sum_i h_i^zm_i$. Suppose for simplicity that  all $h_i^z=0$. For $t_1=\pi/2$ in \eqref{eqelse0} the action of $U_1$ is to flip all the spins $\{m_i\}\rightarrow\{-m_i\}$ while $U_2$ adds a global phase. Clearly {$|\{m_i\}\rangle_z$} cannot be a Floquet eigenstate -- but both Schrodinger-cat like combinations {$|\pm\rangle = (|\{m_i\}\rangle_z\pm|\{-m_i\}\rangle_z$} are $T$-periodic Floquet eigenstates. Such cat states are fragile to perturbations and DTTSB selects one of the pair (at random). Both {$|\{m_i\}\rangle_z$ and $|\{-m_i\}\rangle_z$} evolve periodically at exact resonance with a period 2T. This reasoning can be easily extended to a still soluble case of $t_1=\pi/2$ and $h_i^z\in[0,h^z]$ \cite{ElseFTC}.

The most important feature is that those TTSB solutions are robust against perturbations such as non-zero $h$ or $t_1$ leading to a rotation around $x$ axis by an angle slightly different from $\pi$.  In effect $\langle\sigma_z\rangle$ shows pronounced oscillations with the period of $2T$ (compare Fig.~\ref{nayak1}) revealing DTTSB. The numerical studies use time-evolving block decimation (TEBD) scheme \cite{Vidal2004} for a system of 200 spins and $h=0.3$. It is difficult to separate the effects due to interactions and those due to randomness since the latter is inherent in the model studied.
 
\begin{figure}
 \includegraphics[width=1.0\linewidth]{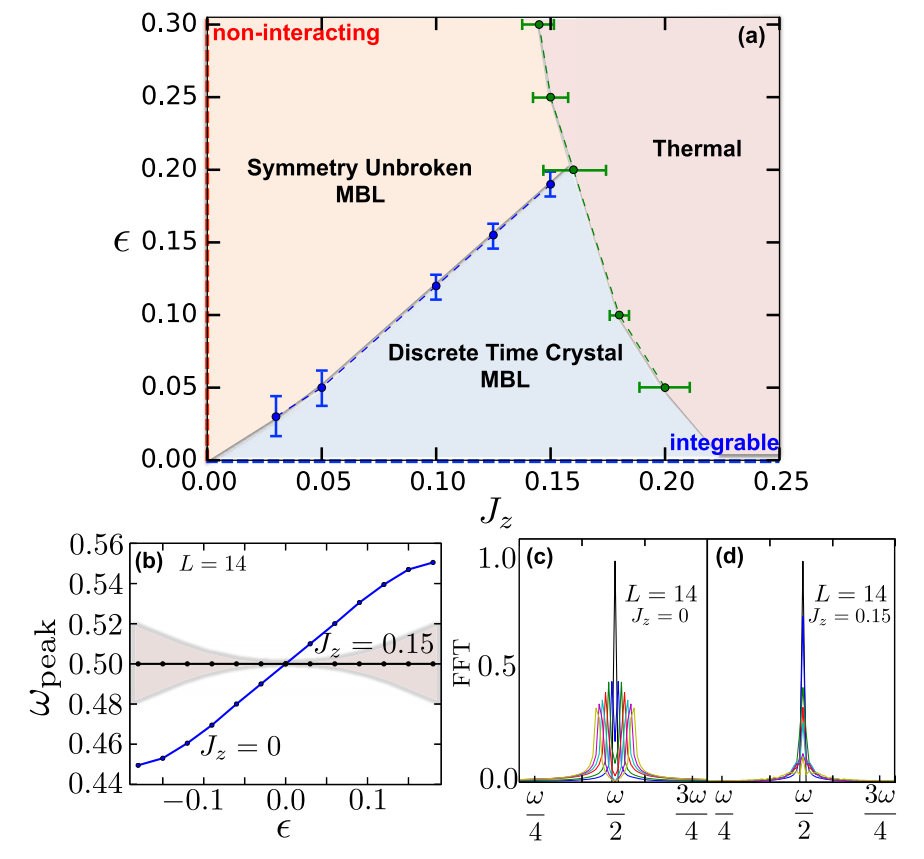}
\caption{Phase space diagram (a) in the plane of $J_z$-$\epsilon$ with $J_z$ being the strength of the interactions while $\epsilon$ the deviation from the perfect spin flip. The discrete (Floquet) time crystal phase appears only in the presence of the interactions and is robust with respect to changes of $\epsilon$. The time crystal phase is identified by the magnitude and variance of the Fourier peak, of the spin ($z$ component) correlation function, at $\omega/2$ as well as the peak position, where $\omega=2\pi/T$. Panel (b) illustrates that in the presence of interactions the peak is locked to $\omega/2$ while
the peak position changes with $\epsilon$ without interactions. Panels (c) and (d) show the Fourier transform shapes without and with the interactions, respectively, corresponding to $\epsilon$ in (b). Reprinted figure with permission from N. Y. Yao, A. C. Potter, I.-D. Potiriche, and A. Vishwanath,, Phys. Rev. Lett, {\bf 118}, 030401 (2017). Copyright (2017) by the American Physical Society \cite{Yao2017} } 
\label{yao1}
\end{figure}

Quite a similar system is studied by Yao and coworkers \cite{Yao2017} who stress the robustness of the observed effect against perturbations. They concentrate on small deviations, say $\epsilon$,  of the ordered phase rotation, i.e. $t_1=\pi/2-\epsilon$ in $U_1$ \eqref{eqelse0} and actually draw a phase diagram in the interactions $J_z$-$\epsilon$ plane -- see Fig.~\ref{yao1}. The system they study is obtained from \eqref{eqelse0}-\eqref{eqelse2} by taking $J_i\in [0.8J_z,1.2J_z]$ and $h_i^x=0$ with $t_2=1$ (we are grateful to Norman Yao for providing us with updated information in the parameters used). The data leading to panel Fig.\ref{yao1}(a) are obtained after averaging over 100 disorder realizations.

\begin{figure}
 \includegraphics[width=1.0\linewidth]{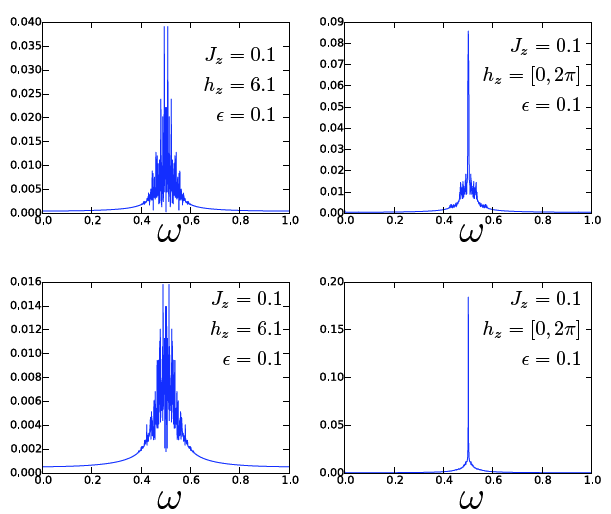}
\caption{Comparison of the Fourier transform of the spin $z$ component correlation function without (left column) and in the presence of (right column) the disorder for the short ranged system \eqref{eqelse0}-\eqref{eqelse2} in the top row and for the long ranged interactions corresponding to the chain of ions \eqref{yaoeq} in the bottom. Figure reprinted from the Supplemental Information to \cite{Yao2017} with the permission of the authors. } 
\label{yao2}
\end{figure}

The main claim of the paper is that there is a phase transition from DTTSB phase to trivial paramagnet 
which is of the Ising type. Importantly, Yao and coworkers \cite{Yao2017} consider not only the abstract spin system but discuss small systems with long range interactions providing quantitative predictions for  controlled ions chain. In that case the unitary evolution over the period $T=t_1+t_2$ is similar to the one given above with interactions between nearest neighbours replaced by the power law decaying long range interactions: $U(T)=U_1(t_1)U_2(t_2)$ with $U_1(t_1)=\exp(-it_1\sum_i \sigma_i^x)$ and 
\be \label{yaoeq}
U_2(t_2)= \exp\left[-it_2\left(\sum_{i\ne j}\frac{J_{ij}}{r_{ij}^\alpha} \sigma_i^z \sigma_j^z +\sum_i h_i^z \sigma_i^z\right)\right].
\ee
Interestingly the DTTSB time crystal behaviour seems to be even more robust with long range interactions as also evident from additional material in supplementary information to that paper. It shows also a difference between a disordered and clean cases showing in the character of the Fourier transform of spin-spin correlation function as reproduced in Fig.~\ref{yao2}. 

\subsection{Experiments}
\begin{figure*}
 \includegraphics[width=1.0\linewidth]{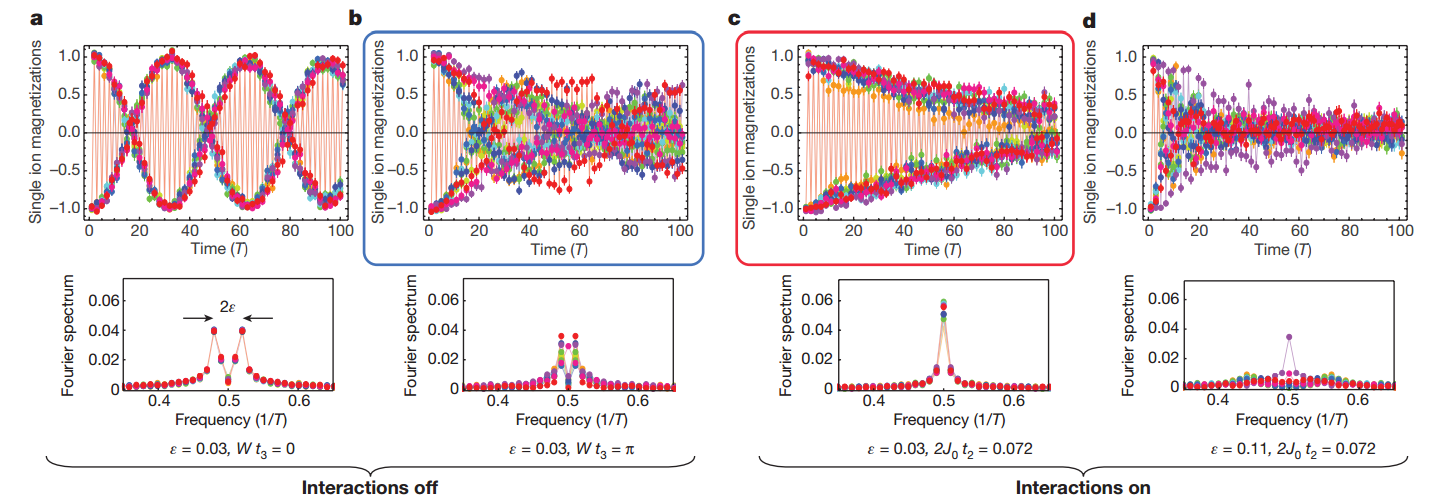}
\caption{Observed magnetization signal as a function of time (in the units of Floquet period $T$) -- top row. Bottom row shows the corresponding Fourier transforms (FTs). In the absence of interactions the spins oscillate with beats corresponding to the perturbation $\epsilon$ from perfect synchronization, those beats may be affected by disorder (panel b). Disorder maximal amplitude is $Wt_3=\pi$ as indicated in the figure. In the presence of interactions (panel c) the self-organization of the system is restored as indicated by a single peak in the FT. As shown in panel d) time crystal survives even in the presence of quite large perturbations. Adapted by permission from Macmillan Publishers Ltd: Nature {\bf 543}, 217-220 (2017)
    doi:10.1038/nature21413 copyright (2017).} 
\label{zhangfig}
\end{figure*}

A detailed proposition of ion chain experiment has proven successful.  The joint effort with experimental team of Chris Monroe resulted in the report of an experimental realization of discrete time-crystals \cite{Zhang2017}. The ion chain consists of 10 $^{171}$Yb$^+$ ions, each providing two sublevels of $^2$S$_{1/2}$ state: $|F=1, m_F=0\rangle$ and $|F =0, m_F=0\rangle$ denoted as effective 1/2-spin states $|\downarrow\ra_z$ and $|\uparrow\ra_z$, respectively. Operations on a single spin are performed using optical Raman transitions. Interactions between spins are due to spin-dependent optical dipole forces. The AC Stark shifts due to off-resonant tightly focused laser beam allow one to introduce programmable disorder addressing each ion. This is accompanied by $\pi/2$ pulses to transfer the disorder
from $\sigma_i^z$ to $\sigma_i^x$. In effect the following Floquet unitary operator is realized:
\bea \label{ion} U(T)&=&\exp(-iH_3t_3)\exp(-iH_2t_2)\exp(-iH_1t_1) \quad \mathrm{with} \nonumber \\
H_1&=&\Omega(1-\epsilon)\sum_i\sigma_i^y, \nonumber \\
H_2&=&\sum_{i,j}J_{ij}\sigma_i^x\sigma_j^x, \\
H_3&=&\sum_i h_i \sigma_i^x. \nonumber
\eea
The Rabi rotation $\Omega=\pi/2t_1$ so for $\epsilon=0$ spins are flipped (as in theory propositions above). Non-zero, controllable $\epsilon$ allows for checking the robustness of the Floquet time crystal formed. The long range interactions $J_{ij}=J_0/|i-j|^\alpha$ with $\alpha=1.5$, site dependent disorder corresponds to $h_i$ uniformly chosen in the interval $[0,W]$ \cite{Zhang2017}.  
While each of the steps in \eqref{ion} looks simple and straightforward the impressive experimental strategy has to be carried out to actually implement \eqref{ion} with the desired precision, the interested reader is advised to consult the original work and the supplementary information \cite{Zhang2017}. 
A single Floquet period lasts $T=t_1+t_2+t_3\approx(15+27+33)\mu s$ being limited by a time needed for sufficient interactions between spins as well as by the fact that disorder $h_i$ is applied to the ions consecutively. 

The system is initiated in the separable state of all spins pointing downwards in the $x$-direction 
$|\Psi\rangle={|\downarrow\downarrow\dots\downarrow\rangle_x}$ 
with $|\downarrow\rangle_x=\frac{1}{\sqrt{2}}(|\downarrow\rangle_z+ |\uparrow\rangle_z)$. 
After evolution with a variable number of Floquet periods the magnetization along $x$ is measured yielding the time correlation function 
\be
\langle\sigma_i^x(t)\rangle=\langle\Psi|\sigma_i^x(t)\sigma_i^x(0)|\Psi\rangle. 
\label{monroe_mag}
\ee
A typical duration of the experiment is about 100 Floquet periods. For optimally chosen $t_1$ [i.e.  for $\epsilon=0$ in \eqref{ion}] and in the absence of the disorder and interactions one expects the magnetization to restore after integer multiples of $2T$ because a single period evolution flips the spins. For $\epsilon\ne 0$, but small, spins rotation reveals beating that is indicated by two peaks in the Fourier transform of the time evolution of the magnetization -- the peaks are located symmetrically around $1/2T$, see Fig.~\ref{zhangfig}(a). Imperfections like the presence of disorder introduce dephasing among the spins, Fig.~\ref{zhangfig}(b). The crucial observation is that the presence of interactions restores the orderly behaviour, the Fourier transform of the correlation function reveals a single peak at half the Floquet frequency, compare Fig.~\ref{zhangfig}(c). The self-organization of the system may be understood as a manifestation of time crystal behaviour. The initial product state in the 
experiment $|\downarrow\downarrow\dots\downarrow\ra_x$ may be thought of as a superposition of two Floquet eigenstates that are Schr\"odinger cat-like states $|\pm\ra=\frac{1}{\sqrt{2}}(|\downarrow\downarrow\dots\downarrow\ra_x\pm|\uparrow\uparrow\dots\uparrow\ra_x)$. The Floquet eigenstates themselves evolve with the period $T$ but they are extremely fragile. Even if the system was prepared in one of the Floquet eigenstates, it would collapse to a product state under infinitesimally weak perturbation and starts evolving with the period $2T$ that indicates spontaneous breaking of the original discrete time translation symmetry.

\begin{figure}
 \includegraphics[width=0.9\linewidth]{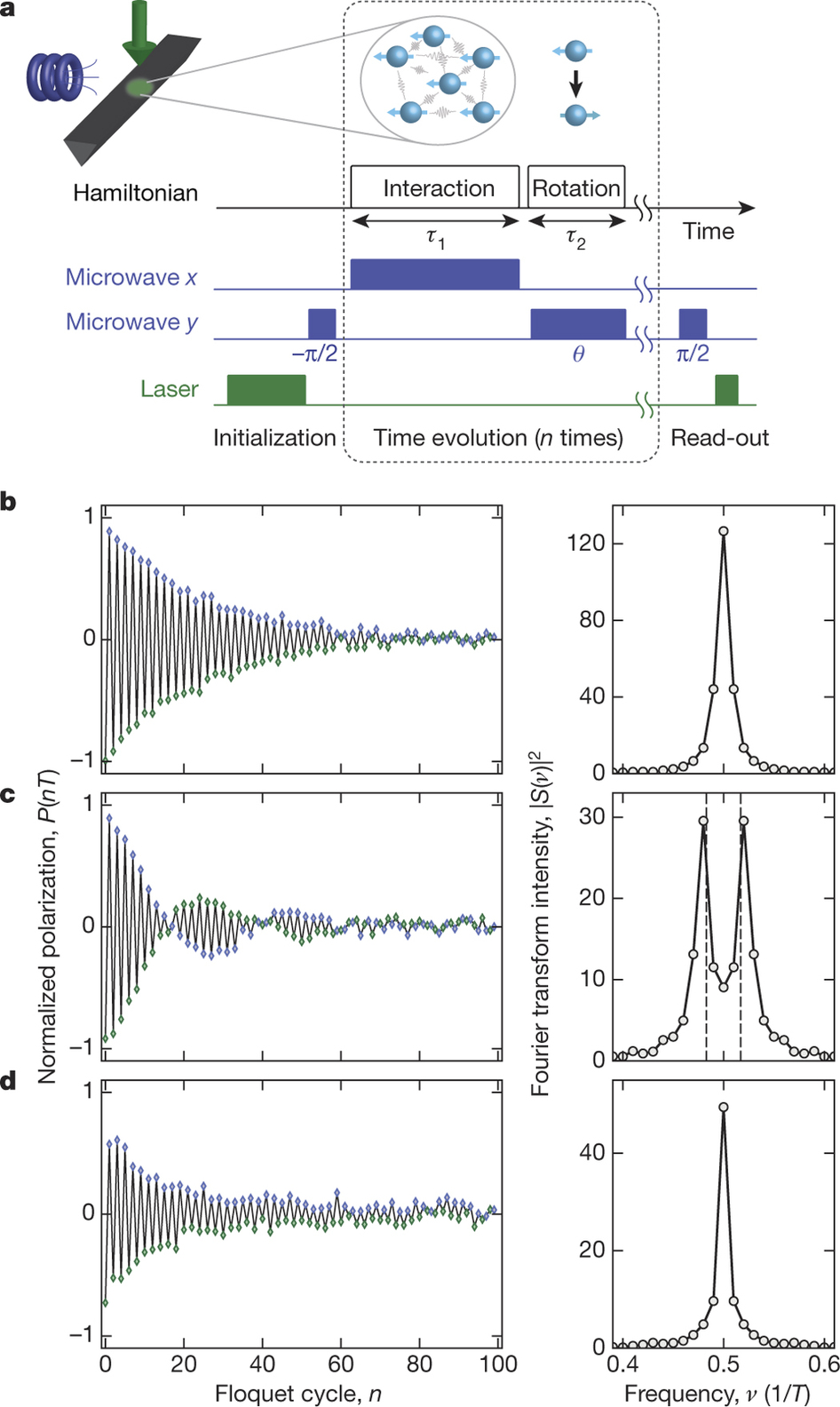}
\caption{Panel (a) shows scheme of the experiment  on nitrogen vacancies. Initially spins are polarized along the positive $x$ direction; during the period $\tau_1$ they evolve in the presence of interactions and microwave field polarization aligned along the $x$ axis; next the spins are rotated around the $y$ axis by angle $\theta\approx \pi$ and the entire procedure is repeated many times. Panels (b)-(d) show observed magnetization (left column) and its Fourier transform (right column). (b) and (c) correspond to short interaction time, the system is sensitive to tuning showing trivial spin-flip signal at resonance when $\theta=\pi$ (b) which becomes modified for rotation detuned $\theta=1.034\pi$ (c). For much longer interaction time, the signal becomes robust and locked at half the frequency corresponding to the period of the drive even for rotation detuned from exact resonance, $\theta=1.034\pi$. Reprinted figure by permission from Macmillan Publishers Ltd: Nature {\bf 543}, 221-225 (2017)
    doi:10.1038/nature21426 copyright (2017).
} 
\label{lukinfig}
\end{figure}

Back to back report in Nature Journal \cite{Choi2017} describes Floquet time-crystalline order observed in a disordered dipolar many body system -- somehow as far from the controlled 10 ions chain as one could imagine. The system studied is an ensemble of nitrogen vacancy spin impurities in diamond. Altogether about 10$^6$ impurities contribute to the signal. Each nitrogen vacancy has unit spin but a two-level system is isolated, by applying an external magnetic field, and it can be manipulated using microwave radiation. In effect the following Hamiltonian is realized:
\bea
H(t)&=&\sum_i \left[\Omega_x(t)\sigma^x_i+   \Omega_y(t)\sigma^y_i+\Delta_i\sigma^z_i\right] \nonumber \\
&+&\sum_{ij}\frac{J_{ij}}{r_{ij}^3}\left(\sigma^x_i\sigma^x_j + \sigma^y_i\sigma^y_j - \sigma^z_i\sigma^z_j\right),
\eea
where $\sigma^\mu_i$ are Pauli spin-1/2 operators, $\Omega_x$ and $\Omega_y$ are Rabi frequencies of microwave drivings, $\Delta_i$ are disorder on-site fields which may be only  qualitatively estimated, $J_{ij}$ are dipolar, direction dependent coefficients with couplings decaying as a third power of the distance between vacancies. 

While the physical system is very different from the one discussed previously, the assumed experimental scheme is a bit similar as shown in the top of Fig.~\ref{lukinfig}. Initially spins are polarized along $x$-axis in the initiation stage. Then during a single period $T$ two strategies are repeated: driving along $x$ with $\Omega_x$ for $\tau_1$ followed by  driving along $y$ with $\Omega_y$ for $\tau_2$ (the Floquet period is $T=\tau_1+\tau_2$). As compared with the former experiment here the disorder and interactions are not switched off/on periodically but are given by a particular system realization.
Almost $\pi$ pulses are realized with $\Omega_y$, $\theta=\Omega_y\tau_2\approx\pi$. A rotation along $x$ with $\tau_1$ serves as an effective ``interaction'' time. For relatively short $\tau_1=92$ns
if $\theta=\pi$ a clear  $2T$-signal is observed in magnetization (or its Fourier transform) -- however deviations from $\pi$ destroy this synchronization, compare Fig.~\ref{lukinfig}(c).
Only when $\tau_1$ time is significantly increased ($\tau_1=989$ns) the Fourier transform peak remains locked at $1/2T$ regardless of the detuning of $\theta$ from the optimal value equal $\pi$. Thus the system self-organizes into a time crystal like behaviour with $2T$ periods. 

It is worth stressing that the same work reports also the $3T$- behavior that may be realized modifying the driving scheme is such a way that after $3T$ the system of spins rotates closely to the initial configuration.

\subsection{Effect of disorder}

Interestingly authors of \cite{Choi2017} stress that their experiment is not performed in the MBL regime, {see also \cite{Zeng2017},} which seems to contradict the claim that MBL is a prerequisite for time crystal behavior \cite{Khemani16,ElseFTC,Keyserlingk16,Yao2017}. Moreover, a recent study of long-range interacting systems \cite{Ho2017} further elaborates on that point -- linking the stability of time crystals in the system studied to quantum critical regime. The authors, using perturbative arguments for their 3D system, suggest, that there is a sharp crossover between a region where the discrete time crystal is stable due to quantum criticality and the regime where discrete time crystal eventually dies. Early, disorder free propositions of discrete time crystals \cite{Sacha2015} obtain their stability via a semiclassical connection to the classical stability of motion in a stable resonant $s:1$ elliptical island. Recently another realizations of clean time crystals in cold atoms have been proposed \
cite{Huang2017}. A series of 
clean quasi one-dimensional models is shown to exhibit Floquet time crystalline behaviour, robust in the strongly interacting regime due to emergent integrals of motion in the dynamical systems studied. It will be interesting to consider the connection between the last two approaches. Last but not least it has been shown that clean Floquet time crystal behaviour may be obtained in the well known kicked Lipkin-Meshkov-Glick model \cite{Russomanno2017}. The stability of the time-crystal phase can be in this model directly analyzed in the limit of infinite size, discussing the properties of the corresponding classical phase space structure. Thus the role played by disorder and MBL in time crystal phenomenon is by no means obvious. There exists the numerical evidence (compare Fig.~\ref{yao2}) that for some spin systems disorder makes time crystal behaviour stronger, still it does not seem to be necessary.

The real area of speculation (or careful analysis) opens when one considers the thermodynamic limit. It is possible that exponentially slow heating may occur, under favorable circumstances in periodically driven many body systems \cite{Abanin2015,Abanin2017}. The resistance of driven systems to expected heating has been even verified experimentally \cite{Bordia2017}. Thus one may consider that before the inevitable thermalization, even in the absence of MBL, for some quite long time one may observe, in a finite system, a time crystal behavior. Such a behavior is called a pre-thermalization and the corresponding time crystal properties have been considered in such a regime \cite{Else17prx}. The authors of diamond experiment \cite{Choi2017} argue that their system cannot be considered as ``pre-thermalizing'' yet they observe the time crystal features in the experiment. It may well be, therefore that for quite large yet finite systems the time crystal behavior remains robust for sufficiently long times that 
other 
mechanisms such as decoherence \cite{Lazarides2017} take over. Clearly more studies of robustness and limitations of time-crystal phenomenon in different settings is {desirable \cite{Moessner2017}.}    
\section{Condensed matter physics in time crystals}

Here we go beyond the demonstration of the time crystal behaviour and present examples in which condensed matter phenomena may be observed in the time domain. The work in this direction is still in its initial stages and one may expect further developments in the near future beyond early proposals revised below.

\subsection{Space periodic systems versus periodically driven systems}

In condensed matter physics it is often assumed that a space crystal is already formed and its properties are analyzed with the help of a space periodic Hamiltonian $H(x+\lambda)=H(x)$. {For ultra-cold
atoms in an optical lattice the space periodicity is imposed
by external laser fields, so a space periodic Hamiltonian arises without the spontaneous symmetry breaking.} Counterparts of space periodic systems but in the time domain are periodically driven systems where $H(t+T)=H(t)$. As mentioned already there is a vast literature on the properties of periodically driven (sometimes called Floquet) systems. We concentrate here on a simple question whether  systems with time periodicity may reveal non-trivial crystalline structures in time.

\begin{figure}
\includegraphics[width=0.49\linewidth]{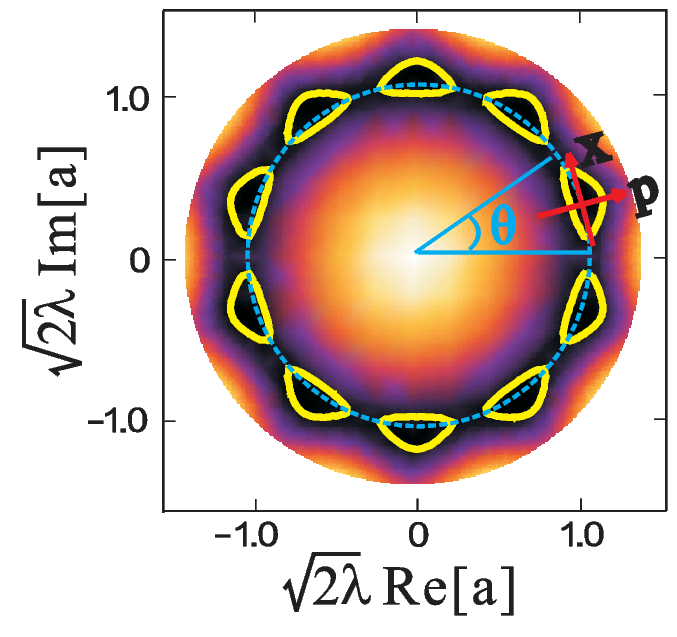}
 \includegraphics[width=0.49\linewidth]{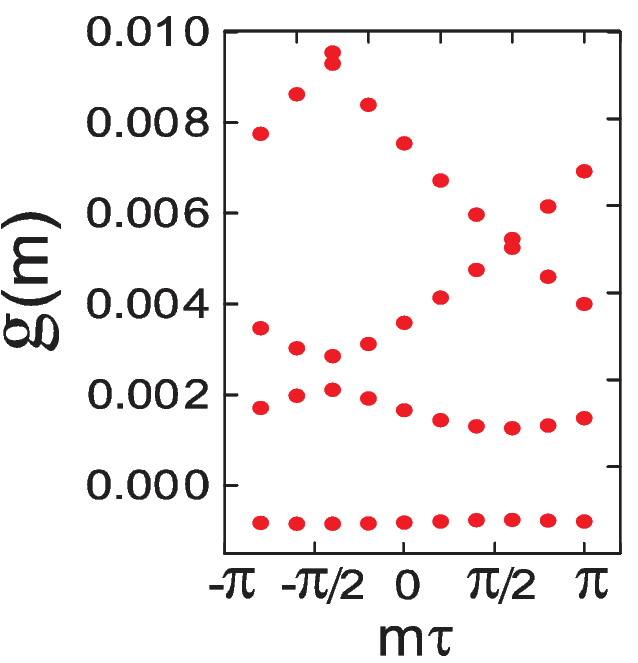}
\caption{Left panel: phase space structure corresponding to a semiclassical version of the Hamiltonian $H_{\rm RWA}$ (\ref{HRWA_Guo}) when the operators are substituted by complex numbers, $\hat a\rightarrow a$ and $\hat a^\dagger\rightarrow a^*$, for $s=10$, {$\mu=4.2\cdot10^{-3}$ and $\lambda\rightarrow 0$, cf. (\ref{hrwa_radial}).} Periodic structure visible in this panel was associated with a phase space crystal in Ref.~\cite{Guo2013}. Right panel: quasi-energy levels $g(m)$ of {(\ref{hrwa_radial}),} that form a band structure, presented in the reduced Brillouin zone, i.e. $m\tau=m2\pi/s\in[-\pi,\pi[$, {for $s=10$, $\mu=3.2\cdot10^{-3}$ and $\lambda=1/205$}. Reprinted and adapted figure with permission from L. Guo, M. Marthaler, and G. Sch\"on, Phys. Rev. Lett, {\bf 111}, 205303 (2013). Copyright (2013) by the American Physical Society \cite{Guo2013}.} 
\label{guo}
\end{figure}

Guo {\it et al.} \cite{Guo2013} have considered a single particle in a perturbed harmonic oscillator potential that is periodically driven,
\be
H=\frac{p^2}{2}+\frac{\Omega^2z^2}{2}+\frac{\nu}{2}z^4+2V_0 z^s\cos\omega t
\label{guo_ham}
\ee
and have shown that the resonant driving, i.e. when $\omega\approx s\Omega$ with integer $s\gg 1$, results in the band structure of the quasi-energy spectrum of the system. This result has been demonstrated within the rotating wave approximation (RWA). That is, applying the unitary transformation $U(t)=e^{i(\omega t/s)\hat a^\dagger \hat a}$, where $\hat a,\ \hat a^\dagger$ are the usual harmonic oscillator annihilation and creation operators, and dropping fast oscillating terms, the effective Hamiltonian becomes
\bea
H_{\rm RWA}&=&\left(\Omega-\frac{\omega}{s}\right)\hat a^\dagger \hat a+\frac{3\nu}{4\Omega^2}\hat a^\dagger\hat a(\hat a^\dagger \hat a+1) 
\cr &&
+\frac{V_0}{(2\Omega)^{s/2}}(\hat a^{\dagger s}+\hat a^{s}).
\label{HRWA_Guo}
\eea
The key observation is that $H_{\rm RWA}$ displays a new discrete symmetry that is not visible in the original Hamiltonian (\ref{guo_ham}), i.e., it commutes with a unitary operator $e^{-i2\pi\hat a^\dagger\hat a/s}$. {It is convenient to introduce} the radial $\hat r$ and the angular $\hat \theta$ operators via $\hat a=e^{-i\hat \theta}\hat r/\sqrt{2\lambda}$ and $\hat a^\dagger=\hat re^{i\hat \theta}/\sqrt{2\lambda}$ where $\lambda=3\nu/(4\Omega^2|\omega/s-\Omega|)$ is an effective Planck constant. { Then, $H_{\rm RWA}=\lambda^{-1}|\frac{\omega}{s}-\Omega|\hat g$ where
\be
\hat g=\frac14(\hat r^2+\lambda-1)^2+\frac{\mu}{2}\left[\left(\hat re^{i\hat\theta}\right)^s+\left(\hat re^{-i\hat\theta}\right)^s\right],
\label{hrwa_radial}
\ee
with 
\be
\mu=\frac{2\lambda V_0}{|\omega/s-\Omega|}\left(\frac{\Omega|\omega/s-\Omega|}{3\nu}\right)^{s/2}.
\ee
Looking at (\ref{hrwa_radial}) it is easy to realize} that $H_{\rm RWA}$ is invariant under the discrete rotation $\hat \theta\rightarrow \hat\theta +2\pi/s$, eigenstates of the system in the angular representation are Bloch-like waves $u_m(\theta)e^{im\theta}$ where $u_m(\theta+2\pi/s)=u_m(\theta)$ and the spectrum splits into bands as illustrated in Fig.\ref{guo}.
{The performed RWA is valid provided $\Omega(\frac{\omega}{s}-\Omega)^2<\frac32 \nu$ what indicates the necessity of the anharmonic term in the Hamiltonian (\ref{guo_ham}) \cite{Guo2013}.}

Guo {\it et  al.} called such a behaviour a phase space crystal. We will see that a phase space crystal is actually a time crystal. That is, any resonantly driven system can be reduced to a solid state-like Hamiltonian by means of the secular approximation approach \cite{Lichtenberg1992,Buchleitner2002} and a periodic structure in the phase space is always reproduced in the time domain \cite{sacha16}.

\begin{figure}
\begin{center}
\resizebox{1.\columnwidth}{!}{\includegraphics{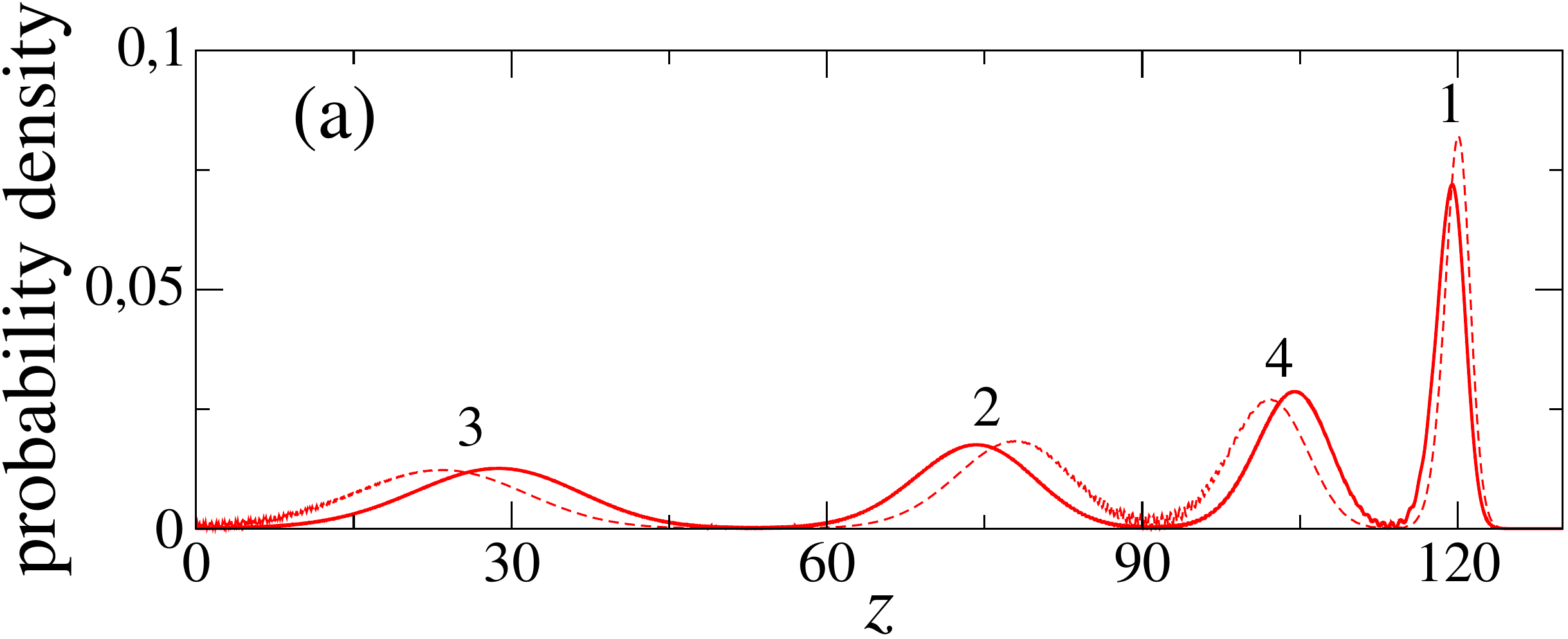}}
\resizebox{1.\columnwidth}{!}{\includegraphics{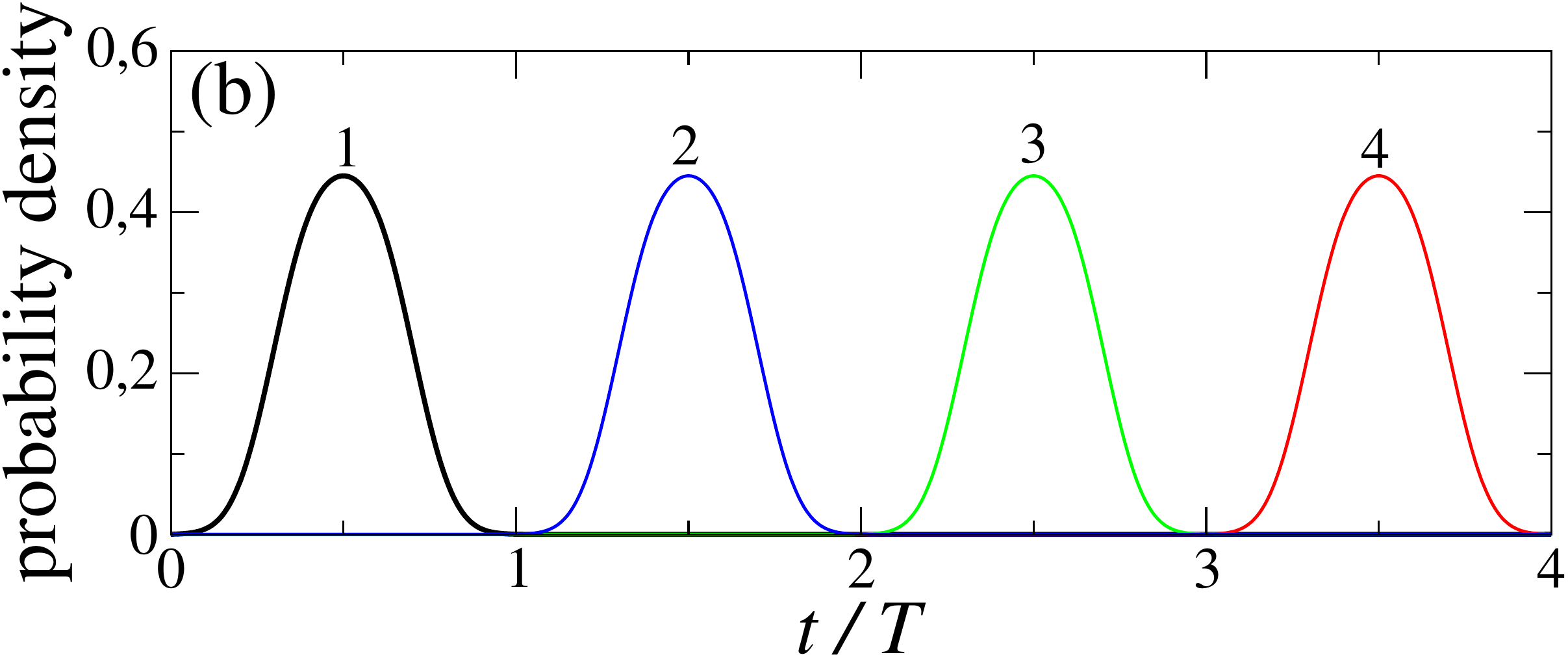}}
\caption{Particle bouncing on an oscillating mirror. Panel (a): plot of a Floquet eigenstate, related to localized wave-packets evolving along $s:1$ classical resonant orbit with $s=4$, in configuration space for two slightly different moments of time, $t=0.25T$ (solid line) and $t=0.3T$ (dashed line). The eigenstate is a superposition of 4 wave-packets that move on the classical orbit. Each wave-packet evolves with the period $4T$ but after $T$ they exchange their positions what makes the whole eigenstate periodic with the period $T$. There are 4 such Floquet eigenstates that are localized on the orbit. Proper superpositions of the eigenstates allows one to extract 4 individual wave-packets, $\phi_j$, that are numbered in (a) and (b). Crystalline structures are not visible in the configuration space, however, they emerge in the time domain. Panel~(b) shows time evolution of the 4 wave-packets whose superpositions form the eigenstates localized on the resonant orbit, at $z=121$ that is close to the turning 
point of a particle in the classical 
description. These wave-packets are analogues of Wannier states corresponding to the lowest energy band of a particle in a spatially periodic potential. Tunneling between wave-packets that are neighbours in the time domain are the leading tunneling processes. Reprinted from \cite{Sacha15a}.
}
\label{fs_sScR15}
\end{center}
\end{figure}

Let us first illustrate this conjecture on the problem of a single particle bouncing on an oscillating mirror in the presence of the gravitational field, see the Hamiltonian (\ref{gpe_s15}). It was said that a suitable choice of the system parameters allows one to realize single particle Floquet eigenstates where localized wave-packets evolve along classical resonant periodic orbits. We have already analyzed the case of the 2:1 resonance. Now let us consider a general $s:1$ resonance \cite{Sacha15a}. With a suitable choice of parameters it is possible to realize Floquet eigenstates that consist of $s$ well localized wave-packets which evolve along $s:1$ resonant orbit each of them with a period $s$ times longer than the period $T$ of the mirror oscillations. They exchange their positions so that the Floquet eigenstates are periodic with the fundamental period $T$, see an example for $s=4$ in Fig.~\ref{fs_sScR15}(a). There are $s$ such Floquet eigenstates $u_1(z,t),\dots,u_s(z,t)$ and they are related to 
quasi-energies $E_1,\dots,E_s$ which are nearly degenerate modulo $\omega/s$. When $s\rightarrow\infty$ these quasi-energies start forming a quasi-energy band. Moreover, one can extract $s$ individual localized wave-packets $\phi_1(z,t),\dots,\phi_s(z,t)$ by a proper superpositions of $s$ Floquet eigenstates. We anticipate that when $s\rightarrow\infty$ these wave-packets become counterparts of Wannier states known in solid state physics that are localized in sites of a spatially periodic potential and are obtained from Bloch eigenstates by a proper superposition
\cite{Kohn1959}.

Now we have all elements in order to show that our system can reveal crystalline structure in the time domain. In Fig.~\ref{fs_sScR15}(a) an example of a Floquet eigenstate is presented in the configuration space for two different moments of time so  one can observe at which direction each of the localized wave-packets is propagating. However, the configuration space is not the domain where we can see crystalline structures. Let us choose a point in the configuration space, e.g. close to the turning point of a classical particle, and plot how the probability density for the detection of a particle at this point changes in time, Fig.~\ref{fs_sScR15}(b). We  can see that each wave-packet arrives at this point periodically and in the same way. That is guaranteed by the Floquet theorem. Thus, the time domain is a proper ``space'' where a crystalline structure emerges. This conjecture becomes clearer when one restricts the description of a particle to the Hilbert subspace spanned by $s$ evolving localized wave-
packets and 
calculates the corresponding quasi-energy of a particle,
 \bea
E&=&\int\limits_0^{s2\pi/\omega}dt\int\limits_0^\infty dz\;\psi^*\left(H(t)-i\partial_t\right)\psi \cr
&\approx& -\frac12\sum\limits_{j=1}^{s}(J_{j}a_{j+1}^*a_j+{\rm c.c.}),
\label{sr_s15}
\eea
where $\psi(z,t)\approx \sum_{j=1}^sa_j\phi_j(z,t)$ was substituted \cite{Sacha15a}. Eq.~(\ref{sr_s15}) is nothing but the kinetic energy term  of the tight-binding model known from solid state physics.  The localized wave-packets, $\phi_j$, play a role of Wannier states and 
\be
J_j=-2\int\limits_0^{s2\pi/\omega}dt\int\limits_0^\infty dz\;\phi_{j+1}^*\left(H(t)-i\partial_t\right)\phi_j
\ee
are tunneling amplitudes. They are related to tunneling of a particle between neighbouring wave-packets (but neighbouring in the time domain) see Fig.~\ref{fs_sScR15}(b). All the tunneling amplitudes have the same absolute values $|J_j|=J$. There are also longer range tunnelings, e.g., between next-to-nearest neighbour wave-packets but they are at least two orders of magnitude weaker in a typical situation and, similarly as in the tight-binding model, can be neglected.

We have demonstrated that the description of a resonant motion of a particle bouncing on an oscillating mirror can be reduced to a tight-binding model known in condensed matter physics \cite{Guo2013,Sacha15a}. We have already mentioned that such an approach can be applied to any resonantly driven dynamical system ranging from ultra-cold atomic gasses to a Rydberg electron perturbed by a microwave field \cite{sacha16}. Indeed, assume that a single particle integrable classical system described by an unperturbed Hamiltonian $H_0$ is driven periodically, i.e. there is another term in a system Hamiltonian, $H=H_0+H_1(t)$, which $H_1(t+T)=H_1(t)$. It is convenient to describe classical motion in action-angle variables of the unperturbed system \cite{Lichtenberg1992}. For example in the case of a 1D system, we can perform canonical transformation from Cartesian position and momentum to canonically conjugate  action $I$ and angle $\theta$ for which $H_0=H_0(I)$. Then, the unperturbed motion can be solved 
immediately, $I=\rm 
constant$ and $\theta=\frac{\partial H_0(I)}{\partial I}t+\theta_0$. The latter equation describes a position of a particle on its trajectory. Now, let us assume that we drive the system harmonically with frequency $\omega$ which is close to a multiple of frequency of an unperturbed motion, 
\be
\omega\approx s\frac{\partial H_0(I_s)}{\partial I_s},
\ee
with an integer $s$. Then, in the moving frame, 
\bea
\Theta&=&\theta-\frac{\omega t}{s}, 
\label{Th_th}
\\  
P&=&I-I_s,
\eea
are slowly varying. Averaging the original Hamiltonian over time we obtain an effective time-independent (secular) Hamiltonian \cite{Lichtenberg1992}
\be
H\approx\frac{P^2}{2m}+V_0\cos(s\Theta),
\label{pendulum_sr15}
\ee
where $m$ and $V_0$ are the effective mass and the amplitude of the effective potential, respectively. Thus, in the moving frame any resonant motion with $s\gg 1$ reduces to a solid state problem of a particle in a space periodic potential Eq.~(\ref{pendulum_sr15}) \cite{Lichtenberg1992,Guo2013,Sacha15a,Guo2016,sacha16}. 

In order to obtain a quantum effective description one can either quantize a classical secular Hamiltonian (\ref{pendulum_sr15}) \cite{Buchleitner2002} or apply the quantum version of the secular approximation from the very beginning \cite{Berman1977}. By restricting to the lowest energy band of a quantum version of (\ref{pendulum_sr15}) one ends up with a tight-binding model of the form of Eq.~(\ref{sr_s15}). Interestingly, it is also possible to consider higher energy bands of the secular Hamiltonian (\ref{pendulum_sr15}) and investigate multi-band physics \cite{Guo2013,Sacha15a,Guo2016,sacha16}.

Note that if a wave-function of a particle in the moving frame, $\psi(\Theta)$, reveals some crystalline structure, in the laboratory frame this wave-function, $\psi(\theta-\omega t/s)$,  will show crystalline properties also versus time $t$ if one fixes position close to a classical orbit because the transformation (\ref{Th_th}) is linear \cite{sacha16}. On the other hand there is no guarantee that such a wave-function will reveal periodic behaviour as a function of a Cartesian position in space for fixed time because canonical transformation between action-angle variables and Cartesian position and momentum is non-linear in general. This is the case for example for a particle bouncing on an oscillating mirror where in the time domain one observes a crystalline structure but does not see it in the configuration space, Fig.~\ref{fs_sScR15}.

\subsection{Anderson localization in the time domain}

Energy eigenstates of a single particle in the presence of a space periodic potential are given by Bloch waves that are extended in space. However, when a potential is not strictly periodic because some disorder is present, eigenstates may become exponentially localized around different points of configuration space due to destructive interference between different multiple scattering paths \cite{Anderson1958}. Regardless of the magnitude of the disorder, provided it is random and decorrelated, Anderson  localization is inevitable in one-dimensional (1D) systems as well as for  time-reversal invariant spinless two dimensional (2D) systems \cite{Abrahams:Scaling:PRL79,MuellerDelande:Houches:2009}.
In the three dimensional (3D) world the situation is more complicated, typically disordered system reveal  the so-called mobility edge. Eigenstates with energies below the mobility edge are localized and those above are extended \cite{Abrahams:Scaling:PRL79,MuellerDelande:Houches:2009}. 

Anderson localization can be also observed in the momentum  space being called then the dynamical localization \cite{Fishman:LocDynAnders:PRL82,Haake,Stockmann,Moore:AtomOpticsRealizationQKR:PRL95,Lemarie:Anderson3D:PRA09}. 
Interestingly, periodically driven single particle systems can reveal Anderson localization in the momentum space despite the classical diffusive behaviour. The diffusion in the phase space is suppressed by quantum interference effects. 

In the present subsection it will be shown that yet another kind of Anderson localization is possible --- Anderson localization in the time domain due to the presence of disorder in time \cite{Sacha15a}, i.e. when the disorder is added on  top of the periodically changing force.  

\subsubsection{Anderson localization in time crystals}
\label{alt}

\begin{figure}
\begin{center}
\resizebox{1.\columnwidth}{!}{\includegraphics{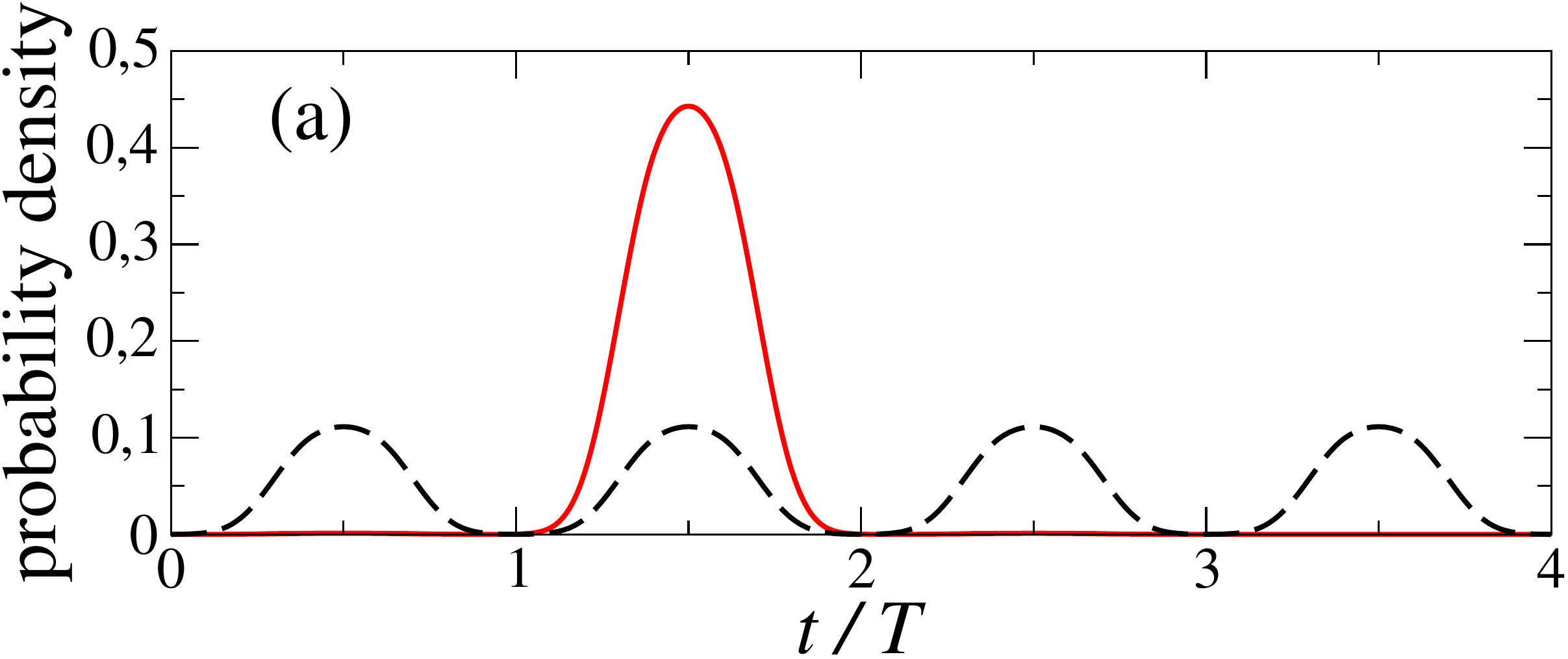}}
\resizebox{1.\columnwidth}{!}{\includegraphics{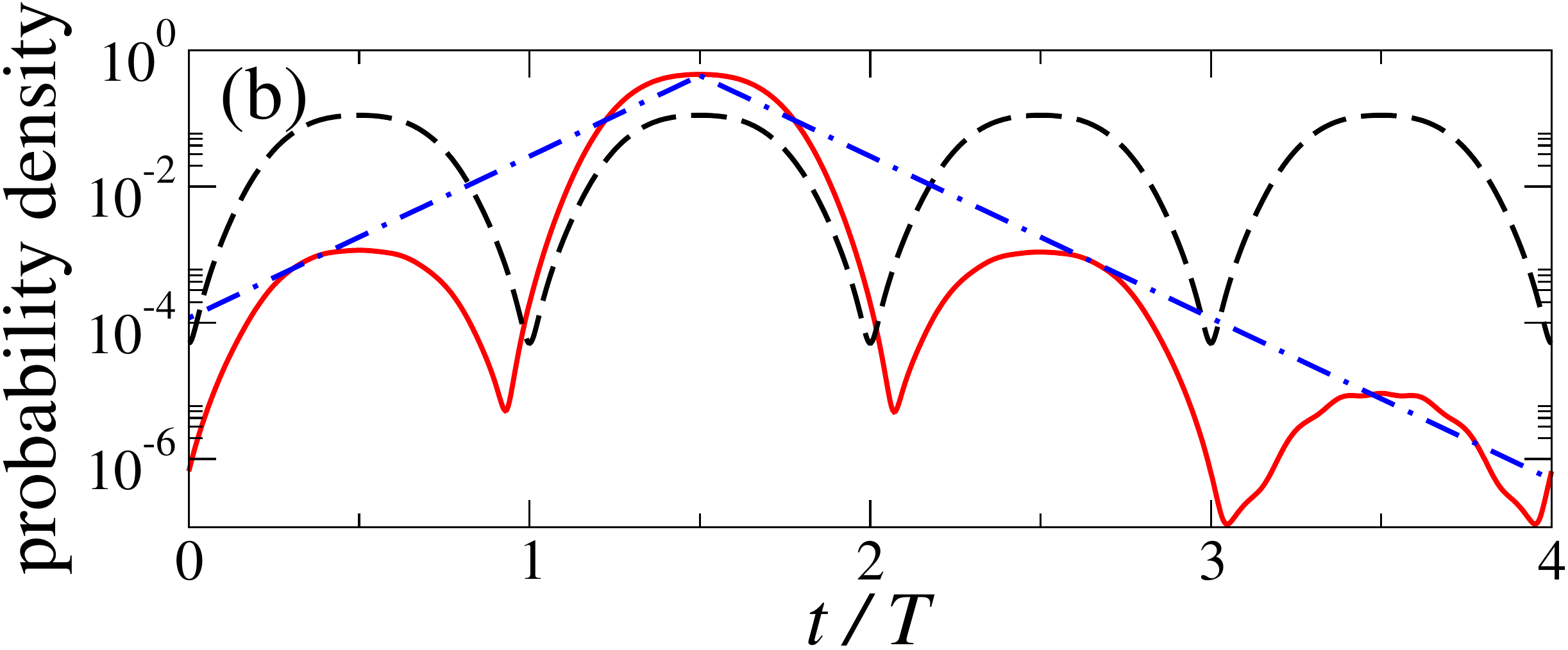}}
\caption{Floquet eigenstate of the system (\ref{disord_s15}), i.e. a particle bouncing on an oscillating mirror (with a period $T$) in the presence of time fluctuating perturbation that is repeated with a period $sT$ with $s=4$. The perturbation is chosen so that $\epsilon_j$'s in (\ref{disord_s15aa}) are random numbers corresponding to a Lorentzian distribution. Then, Eq.~(\ref{disord_s15}) constitutes the Lloyd model of 1D lattice where all eigenstates are Anderson localized and the exact expression for the localization length is known \cite{Haake2001}. From the Lloyd model we know that the eigenstates are superpositions of the wave-packets, $\sum_ja_j\phi_j(z,t)$, with $|a_j|^2\propto e^{-|j_0-j|/l}$ where $l$ is the Anderson localization length and $j_0$ is a number of the wave-packet around which a given eigenstate is localized. The wave-packets $\phi_j$ arrive at a given position $z$ in equidistant intervals in time, thus, the localization length in time is $l_t=lT$. Solid lines in (a) and (b) show one 
of the 4 eigenstates localized on the 4-resonant orbit at $z=121$ versus time in the linear (a) and logarithmic (b) scales. Dash lines present behaviour of the eigenstates in the absence of the disorder in time, i.e. when $H'=0$. Despite the fact that the system is rather small, the characteristic exponential decay of the humps is clearly visible in (b) -- the fitted exponential profile (dash-dotted line) corresponds to $l_t=0.18T$. Reprinted from \cite{Sacha15a}.
}
\label{al_sScR15}
\end{center}
\end{figure}

The simple idea to realize Anderson localization in the time domain is to add random vibrations to a mirror that oscillates with a period $T$ in the bouncing particle problem that has been already described above. It is assumed that a bouncing particle is resonantly driven by a periodically oscillating mirror and $s:1$ resonance condition is fulfilled. Moreover, there is an additional term $H'(t)$ in the system Hamiltonian that introduces a weak temporal disorder --- $H'(t)$ fluctuates in time but fulfills periodic boundary condition on a long time scale, $H'(t+sT)=H'(t)$. It leads to additional terms in the tight-binding energy (\ref{sr_s15}) which now takes the form
 \bea
E&\approx& -\frac12\sum\limits_{j=1}^{s}(J_{j}a_{j+1}^*a_j+{\rm c.c.})+\sum\limits_{j=1}^s\epsilon_j|a_j|^2,
\label{disord_s15}
\eea
where 
\be
\epsilon_j=\int\limits_0^{s2\pi/\omega}dt\int\limits_0^\infty dz\;H'(t)\;|\phi_j|^2,
\label{disord_s15aa}
\ee
are random numbers that can belong to any distribution by a proper engineering of the fluctuating part, $H'(t)$, of the Hamiltonian \cite{Sacha15a}. Energy (\ref{disord_s15}) constitutes actually a 1D Anderson model which possesses exponentially localized eigenstates. In Fig.~\ref{al_sScR15} an example of a localized eigenstate is shown in the case when $\epsilon_j$'s belong to a Lorentzian distribution. 

\begin{figure}
\begin{center}
\resizebox{0.49\columnwidth}{!}{\includegraphics{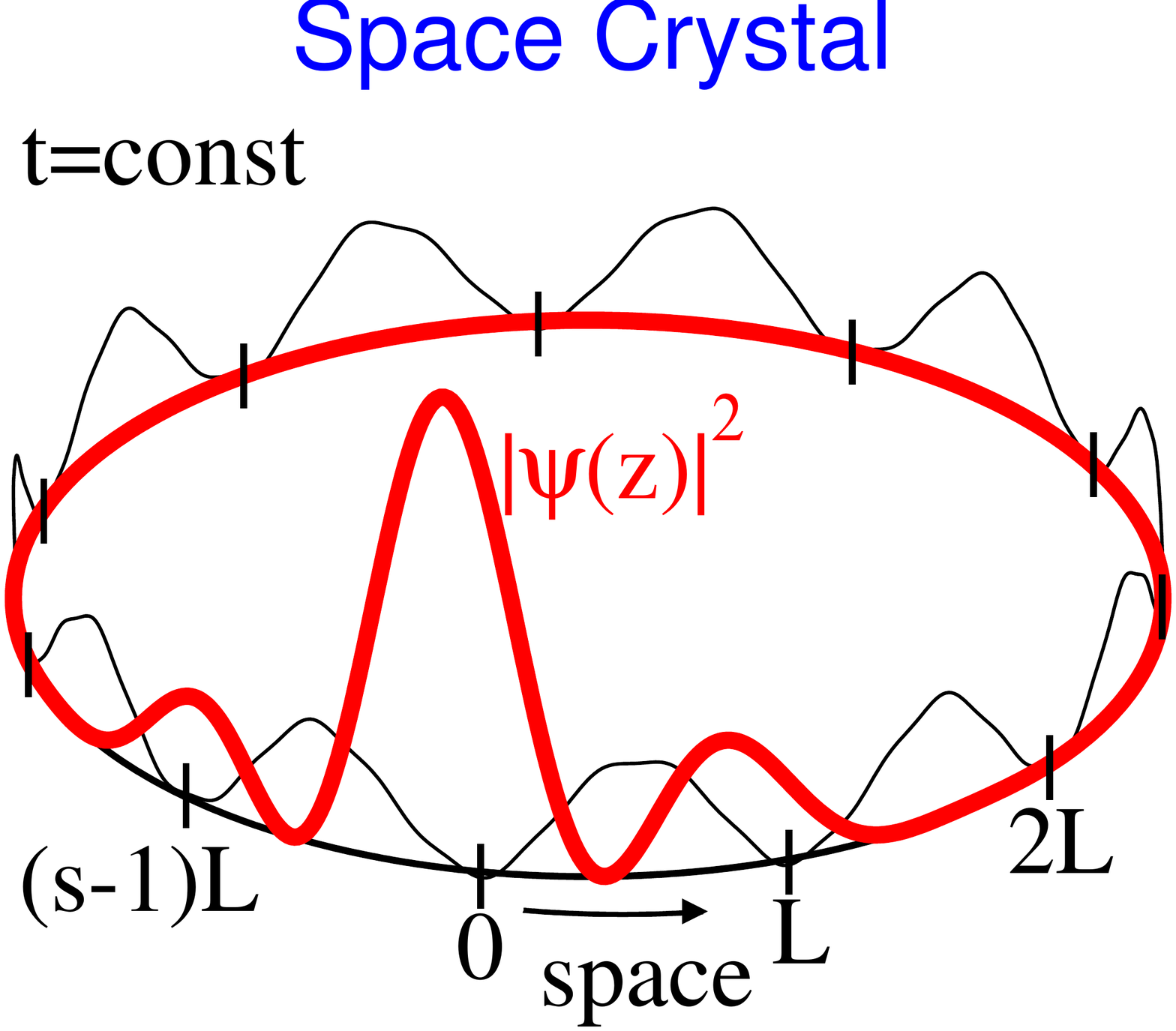}}
\hfill
\resizebox{0.49\columnwidth}{!}{\includegraphics{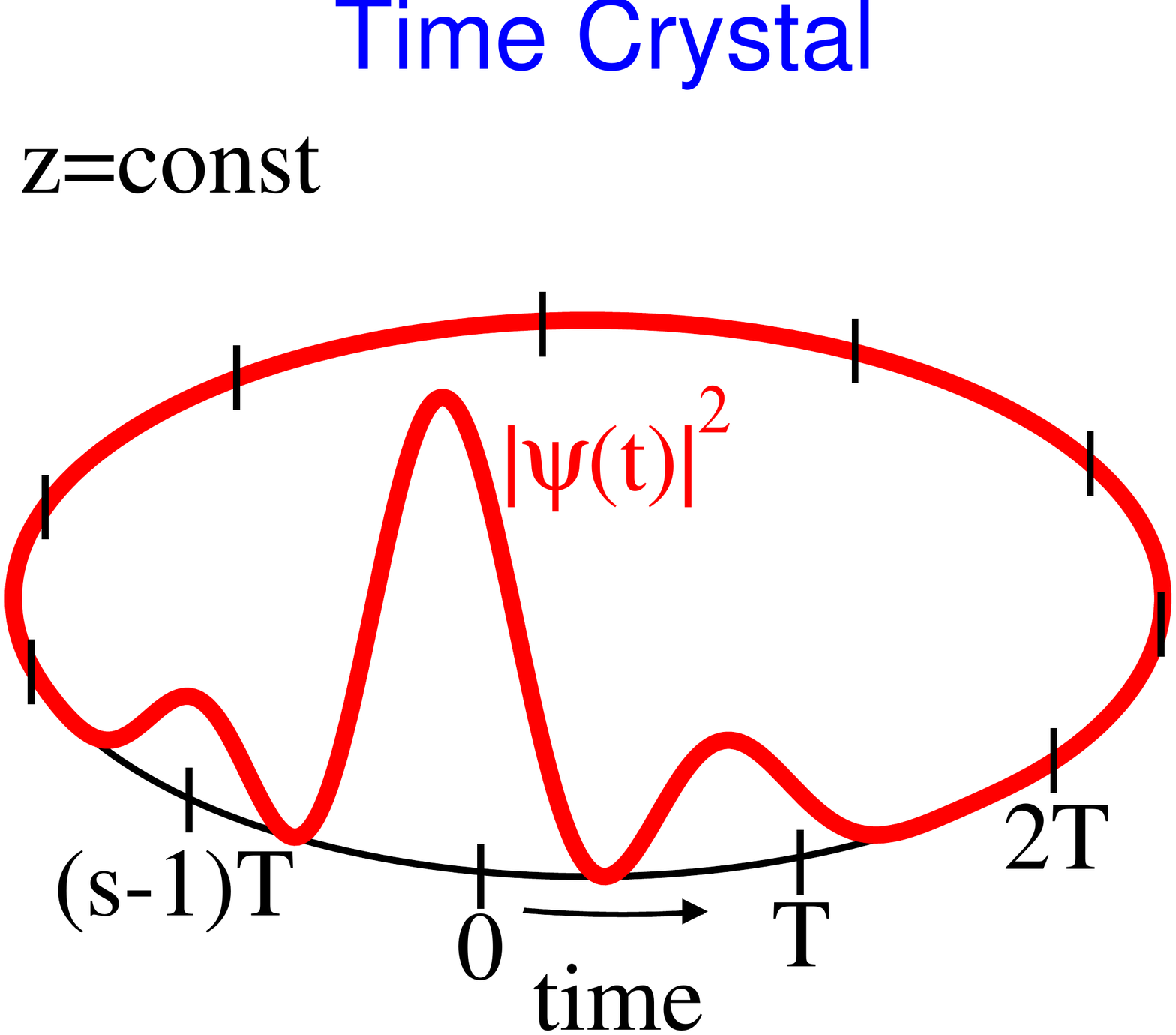}}
\caption{Comparison of Anderson localization in space and time crystals. Left panel shows Anderson localization of a particle in a 1D space crystal with periodic boundary conditions when a disorder is added to a space periodic potential. Due to the periodic boundary conditions when one travels along a ring again and again, one observes periodically a localized wave-function. Right panel illustrates Anderson localization in a time crystal. For a fixed position in configuration space, probability density for a measurement of a particle at this fixed point is localized around a certain time moment. Such a behaviour is repeated periodically with a long period $sT$. 
}
\label{space_time_sScR15}
\end{center}
\end{figure}

The entire Hamiltonian is time periodic with a long period $sT$. It guarantees existence of Floquet eigenstates periodic also with a period $sT$. Anderson localization in the time domain requires that the localization length in time has to be much smaller than $sT$. If a detector is situated close to a classical turning point in the configuration space, it is expected to click with probability that is localized exponentially around a certain moment of time and such a behaviour is repeated periodically with a period $sT$. It is in a complete analogy to Anderson localization in a 1D space crystal with periodic boundary conditions in space (a ring topology). In Fig.~\ref{space_time_sScR15} a comparison of Anderson localization in a time crystal and a 1D space crystal is presented. In the space crystal case when a disorder is present, a particle localizes around a certain space point and moving periodically around the ring, one observes periodically a localized density profile. The larger the ring, the larger 
the 
space crystal. In the time crystal case, probability for a measurement of a particle at fixed space point is exponentially localized in time and this behaviour is repeated periodically with a period $sT$ and the higher $s:1$ resonance the {\it larger} the time crystal.

\subsubsection{Anderson localization in time without crystalline structures}

\begin{figure}
\includegraphics[width=1.\columnwidth]{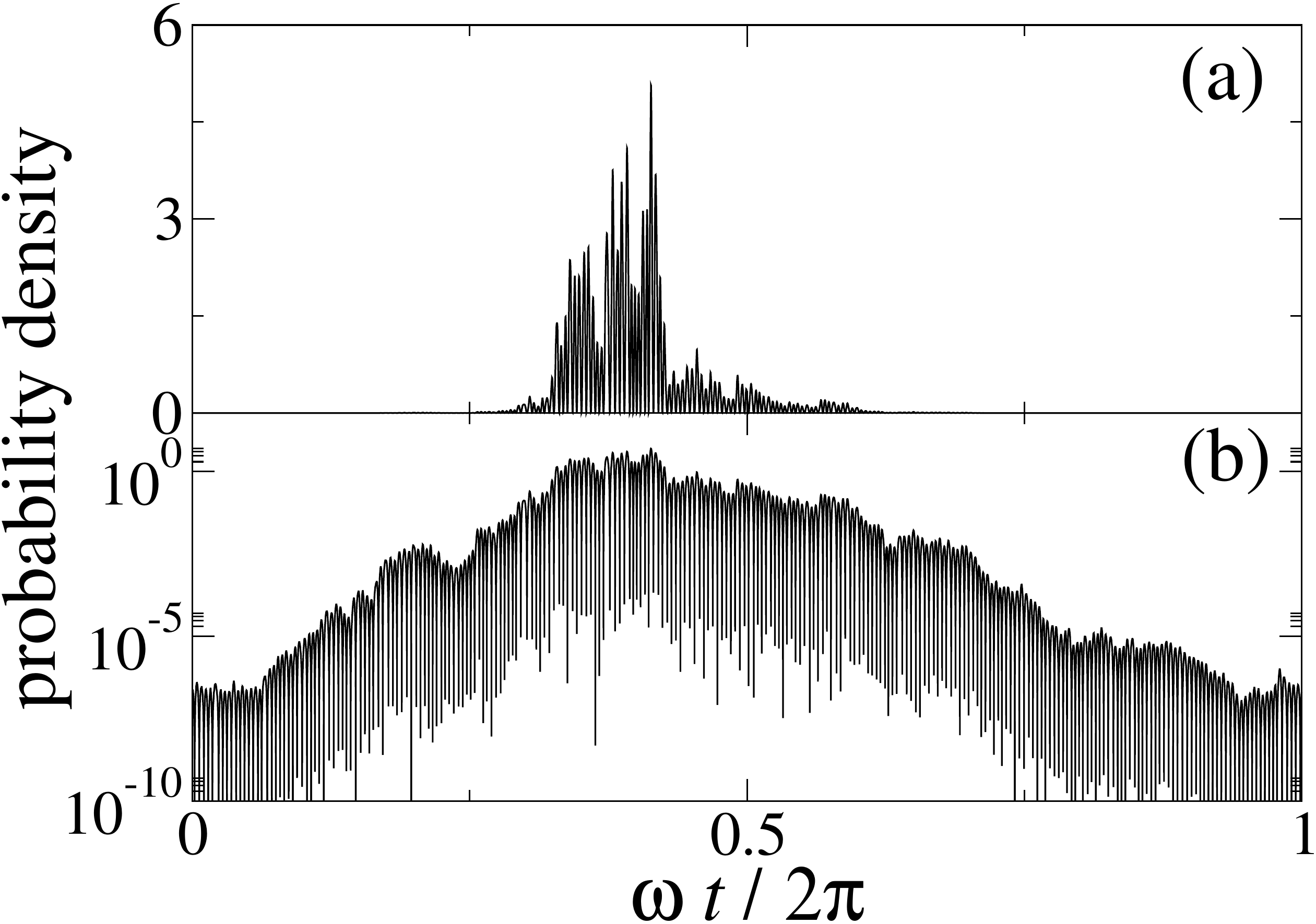}
\caption{Probability density of a Floquet eigenstate corresponding to the Hamiltonian (\ref{h_sd16}) versus time for a fixed position in the laboratory frame. The presented Floquet state reveals Anderson localization in the time domain with localization length of the order of $0.17/\omega$. The upper plot is on a linear scale, the lower plot on a logarithmic scale, showing approximate exponential localization. Reprinted from \cite{sacha16}.}
\label{fig1__PRA_sd16}
\end{figure}

Anderson localization in configuration space takes place when a disorder is present in a space crystal. However, in order to observe Anderson localization it is not necessary to start with a space crystal. One can begin with a purely disordered time-independent potential characterized by a finite correlation length and Anderson localization can be still observed \cite{Kuhn:Speckle:NJP07}. It turns out that an analogous situation takes place in the time domain that will be illustrated with a simple model \cite{sacha16}.

Let us consider a single particle on a ring described by the following classical Hamiltonian,
\be
H=\frac{p^2}{2}+V_0g(\theta)f(t), 
\label{h_sd16}
\ee
where $\theta$ is an angle that denotes a position of a particle on a ring and $V_0$ is amplitude of the perturbation. The perturbation is assumed to be regular in space (for fixed time $t$) and disordered in time (for fixed $\theta$). In the following it is assumed that $g(\theta)=\theta/\pi$ for $\theta\in[-\pi,\pi[$ but because $-\pi$ and $\pi$ correspond to the same point on the ring, $g(\theta)$ has a discontinuity and its Fourier expansion reads,
\be
g(\theta)=\sum\limits_{n\ne 0}g_ne^{in\theta},
\label{g_s15}
\ee 
where $g_n=i(-1)^{n}/(\pi n)$. The function $f(t+2\pi/\omega)=f(t)$ is periodic but between $t=0$ and $2\pi/\omega$ it performs random fluctuations, i.e.
\be
f(t)=\sum\limits_{k\ne 0}f_ke^{ik\omega t},
\ee 
where $f_k=f_{-k}^*$ are independent random variables. Switching to the moving frame, $\Theta=\theta-\omega t$ and $P=p-\omega$, one can see that new position $\Theta$ and momentum $P$ are slowly varying variables if resonant condition is fulfilled, $P\approx 0$. Then, averaging the Hamiltonian over the fast {\it time variable} results in a time-independent effective Hamiltonian,
\bea
H&\approx&\frac{P^2}{2}+U(\Theta), 
\label{heff_sd16}
\\
U(\Theta)&=&V_0\sum\limits_{k\ne 0}g_k f_{-k} e^{ik\Theta}+{\rm constant},
\label{heff_sd16a}
\eea
where a particle moves in the presence of a time-independent disordered potential $U(\Theta)$. This potential can be thought of as the coherent addition of resonant terms between  spatial harmonics of the potential and the corresponding temporal harmonics of the disordered driving amplitude \cite{sacha16}. Statistical properties of Fourier components of $U(\Theta)$ can be engineered by a proper choice of a distribution for $f_k$. The coefficients $g_k$ drops like $1/k$ and in order to deal with the disordered potential characterized by a small correlation length, the drop of $g_k$ has to be compensated by an increase of absolute values of $f_k$. 

Assume that $|g_kf_{-k}|\propto e^{-k^2/(2k_0^2)}$ and ${\rm Arg}(f_{-k})$ are random numbers chosen uniformly in the interval $[0,2\pi[$. With a suitable choice of the amplitude $V_0$ and the correlation length of the disordered potential, which is equal to $\sqrt{2}/k_0$, eigenstates of the Hamiltonian (\ref{heff_sd16}) are Anderson localized even for energies higher than the standard deviation of the disordered potential~\cite{sacha16}. These eigenstates when plotted versus time for a fixed space point in the laboratory frame show up Anderson localization --- an example is presented in Fig.~\ref{fig1__PRA_sd16}.

\begin{figure}
\includegraphics[width=1.\columnwidth]{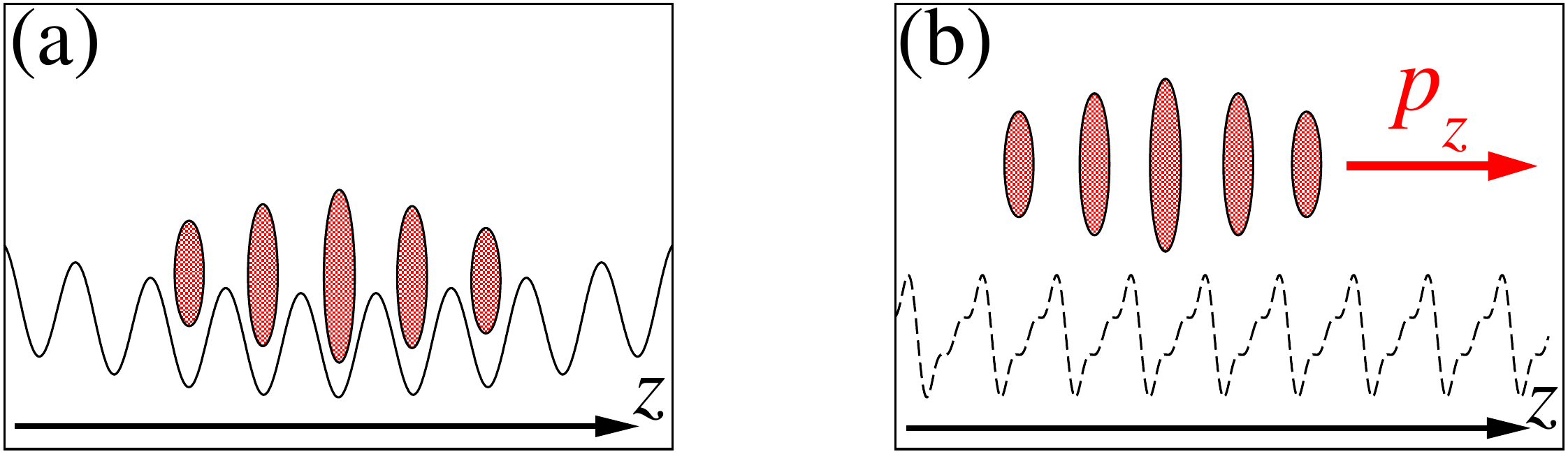}
\caption{ 
Panel (a): schematic plot of the initial stage of the experiment: ultra-cold bosonic atoms are prepared in a strong optical lattice and shallow trapping potentials. It is assumed that the amplitude of the lattice potential is so strong that slices of the atomic cloud are formed that consist of well defined numbers of particles and do not have mutual phase coherence. Panel (b): final stage of the experiment: the optical lattice is turned off, a modulated in time sawtooth-like potential is turned on and atoms are kicked so that their momentum along the $z$ axis fulfills the resonance condition with the modulation frequency. Atoms will fly over the time modulated sawtooth potential and do not spread along $z$ due to the predicted Anderson localization even though their energy is much greater than the amplitude of the sawtooth potential. Reprinted from \cite{delande17}.
}
\label{experiment_dmms17}
\end{figure}

The model (\ref{h_sd16}) can be realized in ultra-cold atomic gases as suggested in \cite{delande17}. Instead of trapping atoms in a ring-shape potential one can prepare a 1D sawtooth-shape potential that is modulated in time in a random manner. It is assumed that ultra-cold atoms are initially prepared in a time-independent optical lattice potential in a Mott insulator phase \cite{Pethick2002} that results in a sequence of independent slices of an atomic cloud, see Fig.~\ref{experiment_dmms17}. Then, the optical lattice is turned off, a modulated sawtooth potential is turned on and atoms are kicked so that their velocity fulfills resonance condition with the modulated sawtooth potential. As a result of Anderson localization in time, the slices of the atomic cloud keep their shape --- they do not spread even though they fly with energy much greater than the amplitude of the modulated sawtooth potential.

Other systems that seem promising for realization of Anderson localization in time are Rydberg atoms perturbed by fluctuating microwave field \cite{Giergiel2017}. A highly excited electron experiences resonant linearly polarized microwave field that performs random fluctuations. An additional static electric field along the microwave polarization axis stabilizes a Kepler orbit an electron is traveling along. It results in an effective Hamiltonian of the form of Eq.~(\ref{heff_sd16}) in the frame moving with an electron because other degrees of freedom of the 3D system are frozen due to the presence of the external fields. One can observe Anderson localization of an electron along a classical Kepler orbit that corresponds to Anderson localization in time if a detector is placed at, e.g., nucleus.

\subsection{Mott insulator in the time domain}

\begin{figure}
\begin{center}
\resizebox{0.45\columnwidth}{!}{\includegraphics{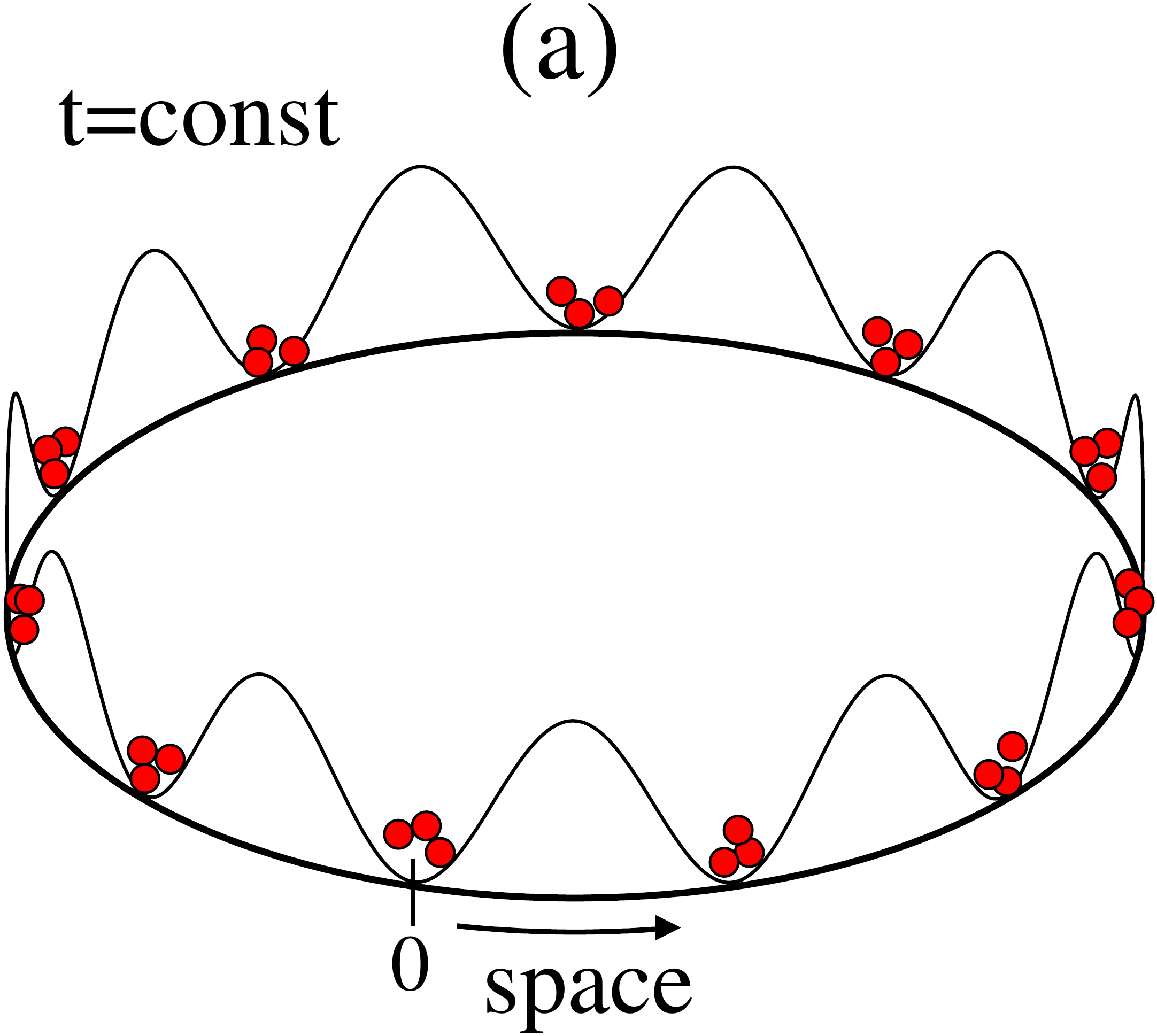}} 
\hfill
\resizebox{0.45\columnwidth}{!}{\includegraphics{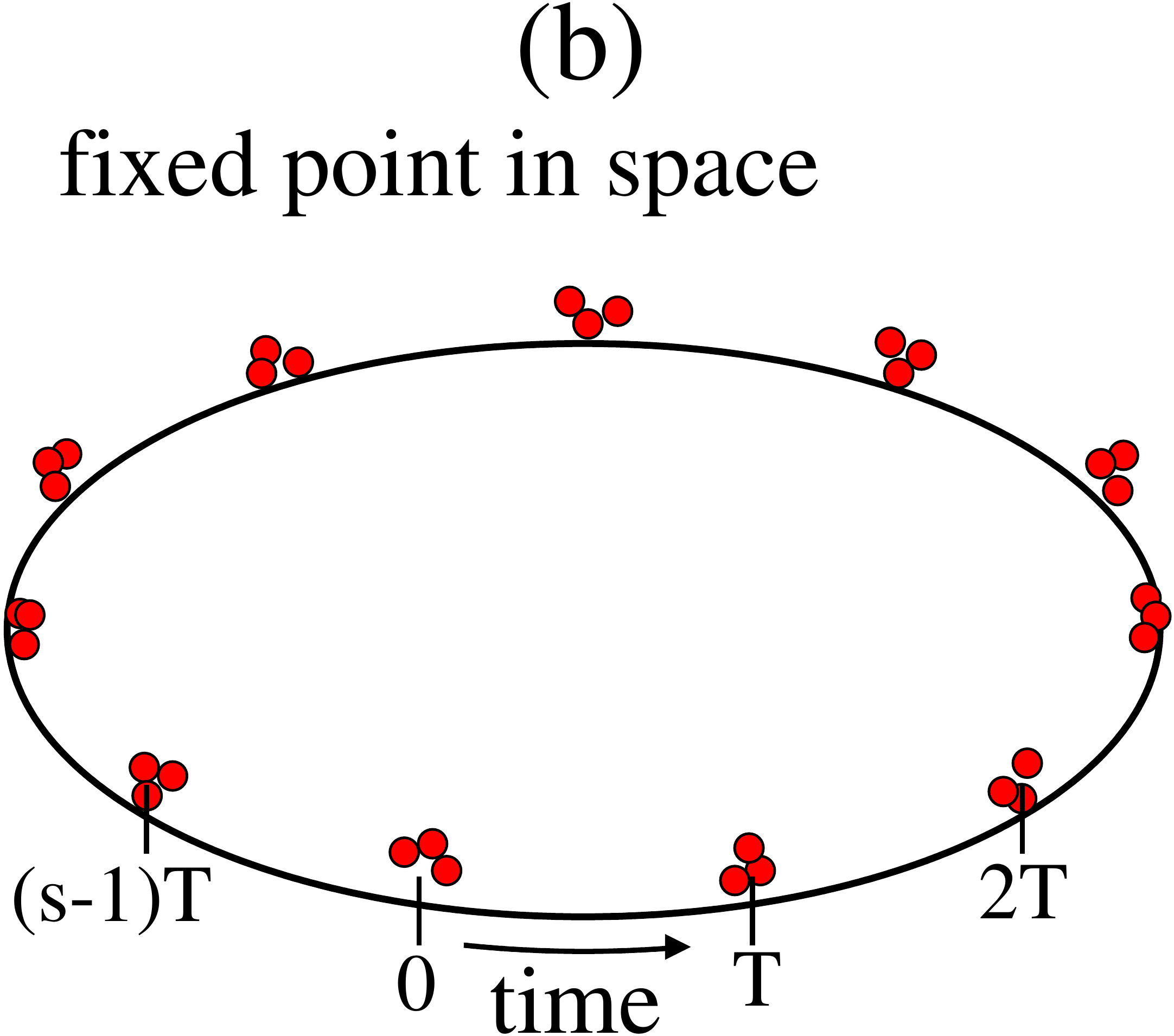}}
\caption{Comparison of Mott insulator phase in the space and time domains. Panel (a) illustrates a space periodic potential in 1D with the periodic boundary conditions, the larger the system, the greater number of potential sites along the ring. Particles in the potential wells illustrate a Mott insulator state where in each potential site there is a well defined number of particles. Panel (b) refers to a crystal structure in the time domain with the time periodic conditions corresponding to the period $sT$. In the Mott insulator regime, at a fixed position, a well defined numbers of particles arrive every time interval $T$. The greater $s$, the larger time crystal is. Reprinted form \cite{Sacha15a}.
}
\label{mott_ScR_s15}
\end{center}
\end{figure}

In the present section we have so far considered single particle systems. However, not only a single atom but a bunch of atoms can bounce on an oscillating mirror \cite{Sacha15a}. Resonant bouncing of a single particle is described by quasi-energy of the form of a tight-binding model, Eq.~(\ref{sr_s15}). In the many-body case a similar simplification can be used if the interaction energy per particle is not much greater than a width of the quasi-energy band of the system (\ref{sr_s15}). 
Close to a $s:1$ resonance a system of ultra-cold bosons with contact interactions may be then described in a restricted Hilbert subspace spanned by Fock states $|n_1,\dots,n_s\ra$, where $n_i$ denotes occupation of a localized wave-packet $\phi_i$ evolving along the $s:1$ resonant orbit. The corresponding many-body Floquet Hamiltonian reads
\bea
H_F&=&\int\limits_0^{s2\pi/\omega}dt\int\limits_0^\infty dz\;\hat\psi\left[H(t)+\frac{g_0}{2}\hat\psi^\dagger\hat\psi-i\partial_t\right]\hat\psi,
\cr
&\approx& -\frac12\sum\limits_{j=1}^{s}(J_{j}\hat a_{j+1}^\dagger \hat a_j+{\rm h.c.})
+\frac12\sum\limits_{j=1}^s U_{ij}\hat a_j^\dagger\hat a_j^\dagger \hat a_j\hat a_j, 
\cr && 
\label{manyHsr_s16}
\eea
where the bosonic field operator $\hat \psi\approx\sum_{j=1}^s\phi_j\hat a_j$ and $|J_{j}|=J$, cf. (\ref{sr_s15}). The coefficients
\be
U_{ij}=g_0\int\limits_0^{s2\pi/\omega}dt\int\limits_0^\infty dz\;|\phi_i|^2\;|\phi_j|^2,
\ee
describe effective interactions between particles that occupy different localized wave-packets. Note that despite the fact the original interactions between particles are contact, the effective interactions can have long range {character \cite{Sacha15a,Anisimovas2015,Eckardt2015,Guo2016a}} because all localized wave-packets pass each other in evolution along a 1D resonant orbit. The on-site interactions are dominating i.e. $|U_{ii}|>|U_{ij}|$ for $i\ne j$.

For sufficiently strong repulsive  ($g_0>0$) contact interactions, when $U_{ii}\gg NJ/s$, the ground state of the system (\ref{manyHsr_s16}) is a single Fock state $|N/s,\dots,N/s\ra$ where well defined numbers of atoms occupy each localized wave-packet, long-time phase coherence is lost, a gap opens between the ground state level and excited levels and consequently a Mott insulator phase is realized in the time domain \cite{Sacha15a}, see schematic picture in Fig.~\ref{mott_ScR_s15}.

\subsection{Many-body localization with temporal disorder}

It is quite natural now to join the results of the previous parts and consider an interacting many-particle system driven resonantly, with 
at the same time added perturbation that fluctuates in time but is repeated with some large period $sT$. That extends the analysis of Section~\ref{alt} on interacting particle systems. Combining with the many body language of the previous Section we are led to a genuine example that may result in a many-body localization induced by temporal disorder. Work in this direction is in progress \cite{dmsz17}.

\subsection{Time crystals with properties of multi-dimensional systems}

It is difficult to predict at which direction, research on time crystals will develop. There are open questions and new ideas are still coming. One of very intriguing topics concerns a possibility of realization of multi-dimensional time crystals. While it is definitely not possible to augment dimensionality of time, one can envisage 1D time crystals but with properties of, e.g., 3D space crystals. 
One proposal was already suggested in the literature that we shortly describe now.

The problem of a single particle on a 1D ring subjected to time-fluctuating perturbation (\ref{h_sd16}) was generalized to the 3D case \cite{delande17}. Consider a particle on a 3D torus described by the following Hamiltonian:
\be
H=\frac{p_\theta^2+p_\psi^2+p_\phi^2}{2}+V_0g(\theta)g(\psi)g(\phi)f_1(t)f_2(t)f_3(t). 
\label{h3d_dms17}
\ee
The function $g(x)$ is the same as in Eq.~(\ref{g_s15}). Time-dependent functions $f_i(t)$ are periodic with periods $2\pi/\omega_i$ but between $t=0$ and $2\pi/\omega_i$ they perform random fluctuations, i.e. $f_i(t)=\sum_{k\ne 0}f_k^{(i)}e^{ik\omega_i t}$ where $f_k^{(i)}=f_{-k}^{(i)*}$ are random numbers. Each degree of freedom is assumed to be in a resonance with one of the time-periodic functions $f_i(t)$ so that in the moving frame where $\Theta=\theta-\omega_1 t$, $\Psi=\psi-\omega_2 t$ and $\Phi=\phi-\omega_3 t$ and for the conjugate momenta $P_\Theta=p_\theta-\omega_1\approx 0$,  $P_\Psi=p_\psi-\omega_2\approx 0$ and $P_\Phi=p_\phi-\omega_3\approx 0$, the effective secular Hamiltonian reads,
\be
H_{\mathrm{eff}}=\frac{P_{\Theta}^2+P_{\Psi}^2+P_\Phi^2}{2}+V_{\mathrm{eff}}(\Theta,\Psi,\Phi),
\label{heff_dms17}
\ee
with $V_{\mathrm{eff}}=V_0h_1(\Theta)h_2(\Psi)h_3(\Phi)$
where 
\be
h_i(x)=\sum_{k\ne 0}g_k f_{-k}^{(i)}e^{ikx},
\label{Eq:defh_dms17}
\ee
provided $\omega_i$ are irrational numbers.
The effective potential in \eqref{heff_dms17} is time-independent and it is a product of three independent disordered potentials along each degree of freedom. From the general theory of Anderson localization in 3D \cite{Abrahams:Scaling:PRL79} one expects the existence of the mobility edge. That is, for properly engineered time fluctuating functions $f_i(t)$ and for a fixed amplitude $V_0$, all eigenstates of (\ref{heff_dms17}) with energies below the mobility edge are Anderson localized and above it are not. In the laboratory frame, when one asks how the probability density for a measurement of a particle at some fixed value of, e.g., $\theta$ changes in time, it turns out it is localized around a certain moment of time provided the relevant Floquet eigenstate corresponds to energy of \eqref{heff_dms17} below the mobility edge \cite{delande17}. 
\section{Conclusions}

In the present article we have reviewed current state of the art of investigations of time crystals originally proposed by Wilczek \cite{Wilczek2012}, i.e. the phenomena that are related to self-organization of quantum many-body systems in time. This kind of self-organization is a truly quantum effect and should be distinguished from classical self-organization phenomena where non-linear oscillators synchronize their motion if a coupling between them is sufficiently strong \cite{Glass1988,Neda2000}. The formation of time crystals is quite analogous to the formation of space crystals. In the space crystal case, a center of mass of a solid state system localizes due to spontaneous breaking of translational symmetry  and the resulting quantum state is no longer invariant under this symmetry. However, a remnant of the symmetry remains, i.e. a discrete space translation symmetry emerges which results in a periodic structure of the probability density for measurements of positions of atoms. 

Time crystals are also related to spontaneous breaking of a translation symmetry but the translation in time. Time-independent systems are invariant under continuous time translation transformation which means that if a system is prepared in an eigenstate, the corresponding probability density does not evolve in time. Frank Wilczek anticipated that it is possible to prepare a many-body system in a state that under an infinitesimally weak perturbation, a system reveals periodic motion \cite{Wilczek2012}. Subsequent studies revealed that such a spontaneous formation of periodic structures in time is not possible if a system is prepared in the ground state or in a thermal equilibrium state (\cite{Bruno2013b,Watanabe2015}).  This is different from the space crystals case where spontaneous formation of crystalline structures is observed in a thermal equilibrium state. 

While the original time crystal proposition turned out to be impossible for realization, Wilczek's vision triggered a new research field and became an inspiration to other scientists. Recently a proposition appeared \cite{Syrwid2017} that avoids the above limitations suggesting to prepare the time crystal using excited states. The main research interest concentrated, however, on a possibility to break discrete time translation symmetry present in periodically driven systems, i.e., on the so called discrete (Floquet) time crystals. This attempt  has been described here  all the way from the birth of the idea to its experimental realizations \cite{Sacha2015,Khemani16,ElseFTC,Yao2017,Zhang2017,Choi2017,Nayak2017}. Discrete time crystals are related to self-reorganization of periodically driven many-body systems. Under certain conditions a many-body system spontaneously switches its period of motion despite the fact that an external driving would like a system to follow its periodic changes. 

Another research direction concerns realization of condensed matter physics in time crystals. Space crystals can be electric conductors or insulators when transport properties in space are studied. It turns out that time crystals are able to possess analogues properties in time. It was already demonstrated that time crystals can be Anderson or Mott insulators \cite{Sacha15a}. While time is a single {\it degree of freedom} and it is not possible to realize multi-dimensional time crystals, time crystals with properties of multi-dimensional space crystals can be imagined and in fact they were already proposed \cite{delande17}. 

There are several possible ways of further developments. One may, for example,  define and classify space-time crystals  exhibiting mixed periodicities in both space and time \cite{Xu2017}. 
One may also pose a question whether time crystal is robust with respect to coupling to an external environment. Time translation symmetry breaking leading to time crystal formation may be considered as an effect of the measurement -- see Section II -- or external perturbation. Still some argue \cite{Lazarides2017} that coupling to the environment must eventually destroy the time crystal while specific example of effective time-crystal creation by decoherence have also been introduced \cite{Nakatsugawa2017}.

Time crystals are also very attractive candidates for quantum simulators that seem to be more flexible than those suggested so far. In the time crystals case, time is an additional degree of freedom, thus, one has an additional {\it knob} to control quantum simulation. 
We believe that a current strong activity in the field of time crystals will reveal novel phenomena that are difficult to discover in condensed matter  systems or simply overlooked so far. Bearing in mind that a time degree of freedom adds an additional dimension, more possibilities for new discoveries are opened.  


\noindent
{\it Note added in proofs.} 
Recently new relevant works appeared in the literature that have not been discussed in the present review article \cite{Gong2017,Iemini2017,Wang2017,Flicker2017,Flicker2017long}.

\section*{Acknowledgments}

We are grateful to Dominique Delande for discussions of various aspects of time crystals.
This work was performed within the  EU  Horizon2020 FET project QUIC (nr. 641122). We acknowledge support of the National Science Centre, Poland via project No.2016/21/B/ST2/01095 (KS) and 2016/21/B/ST2/01086 (JZ).

\vspace{0.5cm}

This is an author-created, un-copyedited version of an article published in Reports on Progress in Physics. IOP Publishing Ltd is not responsible for any errors or omissions in this version of the manuscript or any version
derived from it. The Version of Record is available online at https://doi.org/10.1088/1361-6633/aa8b38.

%

\end{document}